\documentclass[journal]{IEEEtran}

\usepackage{cite}

\usepackage[english]{babel}
\usepackage[utf8]{inputenc}
\usepackage[T1]{fontenc}
\usepackage{dsfont}

\usepackage{amsmath}
\usepackage{amsfonts}
\usepackage{amssymb}
\usepackage{amsthm}
\usepackage{amscd}
\usepackage{bbm}
\usepackage{mathtools,balance}

\usepackage{graphicx}
\usepackage[font=small]{caption}
\usepackage{subcaption}

\usepackage{tikz}
\usetikzlibrary{shadows.blur}
\usetikzlibrary{shapes.symbols}
\usetikzlibrary{decorations.pathreplacing}
\usetikzlibrary{external}
\usetikzlibrary{spy}
\usetikzlibrary{arrows.meta}
\usetikzlibrary{positioning}
\usetikzlibrary{shapes.geometric}
\usetikzlibrary{fit}
\usetikzlibrary{backgrounds}
\usetikzlibrary{calc}

\usepackage{pgfplots}
  \pgfplotsset{compat = 1.17}
  \pgfkeys{/pgf/number format/.cd,1000 sep={\,}}
  \usepgfplotslibrary{external}
  \tikzset{external/system call = {%
    pdflatex \tikzexternalcheckshellescape
      -halt-on-error
      -interaction=batchmode
      -jobname "\image" "\texsource"}}
  \tikzexternaldisable

\usepackage{stfloats}
\usepackage{url}
\usepackage{verbatim}
\usepackage{soul}
\usepackage{enumitem}
\usepackage[linesnumbered, lined, algoruled, algosection, algo2e]{algorithm2e}
\SetKwRepeat{Do}{do}{while}%

\usepackage{algorithm}
\usepackage{etoolbox}
\let\classAND\AND
\let\AND\relax
\usepackage{algorithmic}
\let\AND\classAND
\AtBeginEnvironment{algorithmic}{\let\AND\algoAND}
\floatname{algorithm}{Algorithm}

\newcommand{\trans}{\ensuremath{\mkern-1.5mu\mathsf{T}}}

\newcommand{\frob}{\ensuremath{\operatorname{F}}}

\DeclareMathOperator{\mspan}{span}

\DeclareMathOperator{\rank}{rank}

\DeclareMathOperator{\acc}{acc}

\DeclareMathOperator*{\argmax}{arg\,max}
\DeclareMathOperator*{\argmin}{arg\,min}

\def\tr{{\kern-1pt{\scriptscriptstyle{\top}}\kern-1pt}}

\def\wh#1{\widehat{#1}}
\DeclareMathOperator{\Fone}{F_1}
\DeclareMathOperator{\Ftwo}{F_2}
\DeclareMathOperator{\Prec}{Precision}
\DeclareMathOperator{\Rec}{Recall}
\DeclareMathOperator{\FP}{FP}
\DeclareMathOperator{\FN}{FN}
\DeclareMathOperator{\TP}{TP}

\DeclareMathOperator{\R}{\mathbb{R}}

\def\RNT{\R^{N\times T}}

\def\Rn{\R^n}

\def\Rnn{\R^{n\times n}}
\newcommand{\Ne}{\ensuremath{N_{\operatorname{e}}}}

\newcommand{\hot}[1]{\textcolor{red}{#1}}

\mathcode`\@="8000
{\catcode`\@=\active \gdef@{\mkern1mu}}
\newcommand{\Bzero}{\ensuremath{\mathbf{0}}}

\newcommand{\BC}{\ensuremath{\mathbf{C}}}

\newcommand{\BI}{\ensuremath{\mathbf{I}}}

\newcommand{\BP}{\ensuremath{\mathbf{P}}}

\newcommand{\BR}{\ensuremath{\mathbf{R}}}
\newcommand{\BS}{\ensuremath{\mathbf{S}}}
\newcommand{\BT}{\ensuremath{\mathbf{T}}}
\newcommand{\BU}{\ensuremath{\mathbf{U}}}
\newcommand{\BV}{\ensuremath{\mathbf{V}}}
\newcommand{\BW}{\ensuremath{\mathbf{W}}}
\newcommand{\BX}{\ensuremath{\mathbf{X}}}
\newcommand{\BY}{\ensuremath{\mathbf{Y}}}
\newcommand{\BZ}{\ensuremath{\mathbf{Z}}}
\newcommand{\wtBY}{\ensuremath{\widetilde{\mathbf{Y}}}}

\newcommand{\Be}{\ensuremath{\mathbf{e}}}

\newcommand{\Br}{\ensuremath{\mathbf{r}}}

\newcommand{\Bu}{\ensuremath{\mathbf{u}}}
\newcommand{\Bv}{\ensuremath{\mathbf{v}}}
\newcommand{\Bx}{\ensuremath{\mathbf{x}}}

\newcommand{\Bxhat}{\ensuremath{\wh{\Bx}}}

\newcommand{\BSigma}{\ensuremath{\boldsymbol{\Sigma}}}
\newcommand{\Bmu}{\ensuremath{\boldsymbol{\mu}}}
\newcommand{\Bone}{\ensuremath{\boldsymbol{1}}}

\newcommand{\CB}{\ensuremath{\mathcal{B}}}

\newcommand{\CE}{\ensuremath{\mathcal{E}}}

\newcommand{\CK}{\ensuremath{\mathcal{K}}}

\newcommand{\CO}{\ensuremath{\mathcal{O}}}

\newcommand{\CS}{\ensuremath{\mathcal{S}}}
\newcommand{\CT}{\ensuremath{\mathcal{T}}}

\newcommand{\CX}{\ensuremath{\mathcal{X}}}

\newcommand{\BYk}{\ensuremath{\BY_{K}}}
\newcommand{\BRs}{\ensuremath{\BR_{\CS}}}
\newcommand{\BZs}{\ensuremath{\BZ_{\CS}}}
\newcommand{\BWt}{\ensuremath{\BW^{\CT}}}
\newcommand{\BYs}{\ensuremath{\BY_{\CS}}}
\newcommand{\BYt}{\ensuremath{\BY^{\CT}}}
\newcommand{\BYst}{\ensuremath{\BY^{\CT}_{\CS}}}
\newcommand{\BXst}{\ensuremath{\BX^{\CT}_{\CS}}}
\newcommand{\BCt}{\ensuremath{\BC^{\CT}}}
\newcommand{\etaBs}{\ensuremath{\eta_{\CS}}}

\newcommand{\etaBt}{\ensuremath{\eta_{\CT}}}

\newcommand{\BYo}{\ensuremath{\BY_{\operatorname{obs}}}}

\newcommand{\BYtrain}{\ensuremath{\BY_{\operatorname{trn}}}}
\newcommand{\To}{\ensuremath{T_{\operatorname{obs}}}}
\newcommand{\Ttrain}{\ensuremath{T_{\operatorname{trn}}}}

\newcommand{\BYvoltage}{\ensuremath{\BY_{\operatorname{V}}}}
\newcommand{\BYangle}{\ensuremath{\BY_{\operatorname{A}}}}

\definecolor{matlabblue}{HTML}{0072BD}
\definecolor{matlaborange}{HTML}{D95319}
\definecolor{matlabyellow}{HTML}{EDB120}
\definecolor{matlabpurple}{HTML}{7E2F8E}
\definecolor{matlabgreen}{HTML}{77AC30}
\definecolor{matlablightblue}{HTML}{4DBEEE}
\definecolor{matlabred}{HTML}{A2142F}

\tikzstyle{sline} = [
  solid,
  line width = 1.5pt
]

\pgfplotscreateplotcyclelist{interpApproxPlotlist}{
  {matlabblue, sline, mark=o},
  {matlaborange, sline, mark=+},
  {matlabpurple, sline, mark=+},
  {matlabgreen, sline, mark=+},
  {matlabyellow, sline, mark=+}
}

\pgfplotscreateplotcyclelist{interpColApproxPlotlist}{
  {matlabblue, sline, mark=o},
  {matlaborange, sline, mark=+},
  {matlabpurple, sline, mark=+},
  {matlabyellow, sline, mark=+}
}

\pgfplotscreateplotcyclelist{trainingPlotlist}{
  {black, dashed},
  {matlaborange, sline, mark=+},
  {matlabpurple, sline, mark=+},
  {matlabgreen, sline, mark=+},
  {matlabyellow, sline, mark=+}
}

\pgfplotscreateplotcyclelist{busPlotlist}{
  {black,        sline},
  {matlaborange, sline, dashed},
  {matlabgreen,  sline, dotted},
  {matlaborange, sline},
  {matlabgreen,  sline},
}

\pgfplotscreateplotcyclelist{busPlotlistXie}{
  {black,        sline},
  {matlabgreen,  sline, dashed},
}

\pgfplotscreateplotcyclelist{busPlotlistQdeim}{
  {black,        sline},
  {matlabpurple,  sline, dashed},
}

\newcommand{\plotfontsize}{\footnotesize}

\begin{document}

\title{Interpolatory Approximations of PMU Data: Dimension Reduction and Pilot Selection}

\author{Sean Reiter,~Mark Embree,~Serkan Gugercin,~Vassilis Kekatos,~\IEEEmembership{Senior Member,~IEEE}
        % <-this % stops a space
\thanks{This work was supported by US National Science Foundation grants AMPS-1923221, AMPS-2318800, and EPCN-2500682.}% <-this % stops a space
\thanks{Sean Reiter is with the Courant Institute of Mathematical Sciences, New York University, New York, NY USA 10012 (email: s.reiter@nyu.edu).}
\thanks{Mark Embree and Serkan Gugercin are with the Department of Mathematics, Virginia Tech, Blacksburg, VA USA 24061 (email: embree@vt.edu, gugercin@vt.edu).}
\thanks{Vassilis Kekatos is with the Elmore Family School of Electrical and Computer Engineering, Purdue University, West Lafayette, IN USA 47906 (email: kekatos@purdue.edu).}}

\maketitle

\begin{abstract}
This work investigates the reduction of phasor measurement unit (PMU) data through low-rank matrix approximations.
To reconstruct a PMU data matrix from fewer measurements, we propose the framework of interpolatory matrix decompositions (IDs). 
In contrast to methods relying on principal component analysis or singular value decomposition, IDs recover the complete data matrix using only a few of its rows (PMU datastreams) and/or a few of its columns (snapshots in time).
This row-/column-based compression enables real-time monitoring of power transmission systems using measurements from a smaller subset of pilot datastreams, thereby minimizing communication bandwidth.
The ID perspective gives a rigorous error bound on the quality of the data compression.
We propose selecting the pilot measurements used in an ID via the discrete empirical interpolation method (DEIM), a greedy algorithm that aims to control the error bound. 
This bound yields a computable estimate of the reconstruction error during online operations. 
A violation of this estimate suggests a change in the system's operating conditions and thus serves as a tool for fault detection.
Following a disturbance, DEIM can be used to localize the event source across all buses with high accuracy.
Numerical tests on synthetic PMU data demonstrate DEIM's excellent performance in data compression and validate the proposed DEIM-based fault-detection and localization method.
\end{abstract}


%%%%%%%%%%%%%%%%%%%%%%%%%%%%%%%%%%%%%%%%%%%%%%%%%%%%%%%%%%%%%%%%%%%%%%%%%%%%%%%%
% PAPER CONTENT.                                                               %
%%%%%%%%%%%%%%%%%%%%%%%%%%%%%%%%%%%%%%%%%%%%%%%%%%%%%%%%%%%%%%%%%%%%%%%%%%%%%%%%

% SECTION 1 
%%%%%%%%%%%%%%%%%%%%%%%%%%%%%%%%%%%%%%%%%%%%%%%%%%%%%%%%%%%%%%%%%%%%%%%%%%%%%%%%
\section{Introduction}
\label{sec:intro}
%%%%%%%%%%%%%%%%%%%%%%%%%%%%%%%%%%%%%%%%%%%%%%%%%%%%%%%%%%%%%%%%%%%%%%%%%%%%%%%%
Data-driven methods for real-time power system monitoring have garnered significant attention due to the adoption of phasor measurement units (PMUs) in wide-area monitoring systems (WAMS). PMUs are \emph{in situ} sensor devices that provide global positioning system (GPS)-synchronized phasor readings of nodal voltages, nodal currents, line currents, and their time derivatives, at a rate of 60--120 samples per second. 
These real-time streaming measurements can provide an accurate view of the system's operating condition, enabling operators to monitor network performance, detect disturbances (such as line trips or outages), and initiate corrective measures.
However, data accumulation presents a significant roadblock to real-time operational benefits: e.g., a network of 100~PMUs, each with a sampling rate of 120~Hz, generates 200~gigabytes of data per day~\cite{KliATK10, GadBBC16}. 
Moreover, a large communication bandwidth is required to transmit 
PMU data to control centers~\cite{DasS13, Das16, XieCK14}.
Thus, reliably managing and reducing the scale of streaming PMU data becomes an essential research area.

PMU time series data can be organized in a matrix form: each row corresponds to a single PMU measurement stream, e.g., voltages at a particular bus, and each column contains a snapshot in time.
It is well-documented in theory and industry practice that matrices of PMU data exhibit an approximate low-rank structure under both normal and abnormal conditions.
This low-rank phenomenon has been attributed to the spatial-temporal correlations in PMU datastreams;
see, e.g.,~\cite{DahKM12, XieCK14, WangEtal21, LiWC18}.
Data-driven methods exploiting such dependencies in PMU data have been successfully applied to grid monitoring tasks, including detection and localization of disturbance events.
Low-rank representations of PMU data are typically computed using methods that decompose the data into orthogonal components, e.g., the singular value decomposition (SVD)~\cite{GolV12} or the closely related principal component analysis (PCA)~\cite{Jol02}.
Event detection algorithms based on PCA are proposed in~\cite{WanZZ11, Liuetal15, RafLLM16}.
Recognizing the underlying low dimension of PMU data, the authors in~\cite{XieCK14} propose to monitor a network using fewer pilot PMUs.
In~\cite{LiWC18}, low-dimensional subspaces derived from the SVD and subspace comparison metrics are used to identify, detect, and localize events in real time.
The papers~\cite{Gaoetal15, Haoetal18} develop matrix completion-based methods for event detection that use the SVD~\cite{Gaoetal15} and arrange PMU data into low-rank Hankel matrices~\cite{Haoetal18}.
The paper~\cite{KonFYY21} seeks to express a PMU matrix as a sum of a low-rank matrix, a noise matrix, and a row-sparse matrix that captures abnormal network behavior; this structure is leveraged to detect events.
See~\cite{WangEtal21} for a survey of low-rank methods in WAMS.

PCA/SVD provide \emph{optimal} low-rank approximations to a complete PMU matrix by blending information from \emph{all} rows and columns of the PMU matrix, and thus require large amounts of PMU data to be communicated across the network before compression can be applied at a central hub, such as a phasor data concentrator (PDC).\ \ Thus, PCA/SVD can be ill-suited for time-sensitive and bandwidth-limited tasks. Moreover, applications such as post-event analysis
do not explicitly seek an optimal reconstruction of the PMU matrix, but rather aim to reveal a small subset of rows and columns that capture its low-dimensional structure and correspond to points of interest in the grid's operating history. 

As an alternative, we explore \emph{interpolatory matrix decompositions} (IDs)~\cite{MahD09, SorE16, LWMRT07, DonM23, Ste99}
for the real-time dimension reduction (compression) of PMU data to enable fast and reliable methods for WAMS.\ \ 
In contrast to PCA/SVD, which use all available measurements, IDs aim to reconstruct the complete PMU data matrix using measurements collected from only a small number of PMU datastreams, which we refer to as \emph{pilot streams}~\cite{XieCK14}, and/or from downsampled time snapshots~\cite{Liuetal19, DasS13, Das16}, which we call \emph{pilot snapshots}.
While suboptimal in approximation quality, an interpolatory decomposition is more suitable for \emph{real-time} applications in WAMS.\ \ 
For instance, a reduced set of pilot streams can be monitored during real-time operations and used to approximately represent any non-pilot datastream via an ID, thereby significantly reducing the dimension of streaming PMU data.

\emph{Pilot selection} is the task of selecting a reduced number of pilots offline to monitor during online operations for the purpose of minimizing communication bandwidth.
The quality of the online dimension reduction achieved via an interpolatory approximation hinges on the choice of pilots, which are identified during an offline training phase.
We propose using the \emph{discrete empirical interpolation method} (DEIM)~\cite{SorE16, ChaS10, Baretal04, DrmG16, Hen24} for adaptively performing this pilot selection.
DEIM is a greedy algorithm that aims to minimize a computable upper bound on the interpolatory approximation error.
This bound can certify whether a given collection of pilot streams or snapshots truly captures the low-rank character of the data, and can be leveraged for operational uses, such as online error estimation and disturbance event detection.

Combining the framework of IDs with DEIM, we propose an offline–online framework for real-time event monitoring using a reduced number of pilot datastreams.
Offline, DEIM parses ambient data from all PMUs to select a minimal set of pilot streams until the interpolatory error bound meets a user-specified tolerance.
Online, only data from these pilots are communicated to the control room. 
PMU data from the non-pilot buses are readily reconstructed from the pilot datastreams via an ID.
Our work especially builds on the fundamental contribution of~\cite{XieCK14},
whose authors pursue a similar goal of dimension reduction for real-time event detection.
We observe that their method is actually a form of ID, and thus the error bound~\eqref{eq:oneSidedInterpErrorBound} applies to it. Our approach differs in its general framework based on IDs and in the way it selects pilots.
By using DEIM, we seek to control the error bound, so much so that~\eqref{eq:oneSidedInterpErrorBound} serves as a viable error estimator during online operations, and provides a lightweight tripwire for detecting disturbances.

\emph{Contributions.}
The key contributions of this work are:
\begin{itemize}
\item Introducing IDs as a unified framework for compressing PMU data from fewer rows and/or columns, enabling the use of a rigorous error bound
for certifying IDs for PMUs;
\item Proposing the use of the discrete empirical interpolation method (DEIM) for selecting pilot streams and snapshots;
\item Adapting the interpolatory error bound into an indicator for event detection during online monitoring; 
\item Applying DEIM to post-event data to locate faults.
\end{itemize}

\emph{Organization.}
We review the basics of IDs in the context of PMU data reduction in Section~\ref{sec:lowRankPMU},  and describe how these low-rank methods can be used for the real-time reconstruction of streaming PMU data. 
Section~\ref{sec:DEIM} introduces DEIM for selecting the pilots that determine the ID.\ \  Building upon~\cite{XieCK14},
in Section~\ref{sec:eventMonitoring} we describe an ID-DEIM framework for data-driven monitoring of power systems using a reduced set of pilot streams.
Numerical tests using synthetically generated PMU data illustrate and validate the proposed methodology.

\emph{Notation.}
Bold lowercase and uppercase letters $\Bx\in\Rn$ and $\BX\in\R^{n_1\times n_2}$ denote vectors and matrices.
We use MATLAB notation to index:
the $(i,j)$-th entry of $\BX$ is denoted $\BX(i,j)\in\R$; the $i$-th row of $\BX$ is  $\BX(i,\,:)\in\R^{1\times n_2}$; the $j$-th column of $\BX$ is  $\BX(:,\,j)\in\R^{n_1}$ and occasionally $\Bx_j\in\R^{n_1}$.
Given a set of indices $\CK=\{k_1,\ldots, k_m\}$, let $\BX(\,:,\,\CK)\in\R^{n_1\times m}$ and $\BX(\CK,\,:\,) \ \in\R^{m\times n_2}$ denote the columns and rows of $\BX$ indexed by $\CK$, and $m=|\CK|$ denote the cardinality of $\CK$.\ \ 
With $\BI_n\in\Rnn$, $\Be_i\in\Rn$, and $\Bone_n\in\Rn$ we denote the $n\times n$ identity, the $i$-th canonical unit vector ($\Be_i$ is one in entry $i$ and zero elsewhere), and the vector of all ones.
We use $\|\cdot\|_2$ and $\|\cdot\|_{\frob}$ to denote the spectral and Frobenius norms of a matrix, and $\cdot^{\trans}$ to denote the vector/matrix transpose.

% SECTION 2
%%%%%%%%%%%%%%%%%%%%%%%%%%%%%%%%%%%%%%%%%%%%%%%%%%%%%%%%%%%%%%%%%%%%%%%%%%%%%%%%
\section{Low-Rank Approximation of PMU Matrices} \label{sec:lowRankPMU}
%%%%%%%%%%%%%%%%%%%%%%%%%%%%%%%%%%%%%%%%%%%%%%%%%%%%%%%%%%%%%%%%%%%%%%%%%%%%%%%%
Suppose a system operator collects data at $T$ discrete time instances from $N$ PMU datastreams. 
To simplify the exposition, assume each measured bus is instrumented with a single PMU.\ \ 
We assume that each row of a matrix $\BY\in\R^{N\times T}$ containing PMU time series data corresponds to a single measured quantity, e.g., each row contains voltage magnitudes from a particular bus. Multiple measured quantities are handled by organizing the data into distinct matrices and processing each matrix separately.
We envision that $\BY$ is typically wide ($T>N$) due to the high sampling rate of PMUs, but this is not required; our discussion applies to $\BY$ of arbitrary dimension.

The underlying dimension of PMU data has been considered from a variety of perspectives; see, e.g.~\cite{DahKM12, DasS13, XieCK14, Das16, WangEtal21, LiWC18, Haoetal18}. 
Matrices of PMU data typically exhibit an underlying low-rank structure, regardless of whether the data are collected during ambient or irregular operating conditions. 
As a result, the underlying dimension (rank) of $\BY$ can be reduced 
by retaining only its dominant components computed via PCA~\cite{Jol02} or the SVD; see Section~2.4 of~\cite{GolV12}.
The low-rank factors can be stored more efficiently, and subsequent computations and analyses on the reconstructed data can be expedited.

%%%%%%%%%%%%%%%%%%%%%%%%%%%%%%%%%%%%%%%%%%%%%%%%%%%%%%%%%%%%%%%%%%%%%%%%%%%%%%%%
\subsection{Approximations of PMU Data via the SVD/PCA}
\label{ss:SVD}
%%%%%%%%%%%%%%%%%%%%%%%%%%%%%%%%%%%%%%%%%%%%%%%%%%%%%%%%%%%%%%%%%%%%%%%%%%%%%%%%
We briefly review how one can obtain a low-rank matrix approximating $\BY$ by truncating the trailing components of its SVD. 
Define $R = \rank(\BY)\le \min\{N,T\}$. The SVD of $\BY$ is 
\begin{equation}\label{eq:svd}
\BY = \BU\BSigma\BV^{\trans} = \sum_{k=1}^R \sigma_k^{} \Bu_k^{}\Bv_k^{\trans}
\end{equation}
where the diagonal matrix $\BSigma\in\R^{R\times R}$ carries the singular values $\sigma_1\geq \sigma_2\geq \cdots\geq \sigma_R>0$, and matrices $\BU\in\R^{N\times R}$ and $\BV\in\R^{T\times R}$ have orthonormal columns $\Bu_1,\ldots,\Bu_R$ and $\Bv_1,\ldots,\Bv_R$ called the left and right singular vectors.

Because real-world PMU data are corrupted by noise, $\BY$ typically has full rank $R = \min(N, T)$. Nonetheless, it can be approximated well by a matrix of lower rank $K\ll R$ obtained from the leading terms in the SVD.\ \ 
By the Eckart--Young--Mirsky theorem (see, e.g., Theorem.~2.4.8 of \cite{GolV12}), an optimal rank $K<R$ approximation to $\BY$ in the spectral and Frobenius norms is 
\begin{equation}
\label{eq:svdK}
\BYk\coloneqq\BU_K^{}\BSigma_K^{}\BV_K^{\trans} = \sum_{k=1}^K \sigma_k^{}\Bu_k^{}\Bv_k^{\trans}
\end{equation}
obtained by summing the leading rank-one components $\sigma_k^{}\Bu_k^{}\Bv_k^{\trans}$ corresponding to the largest $K$ singular values. Here $\BU_K \in \R^{N\times K} ,\BSigma_K \in \R^{K\times K},$ and $\BV_K \in \R^{T\times K}$ are the sub\-matrices of $\BU,\BSigma,$ and $\BV$ associated with those leading $K$ singular values. 
The matrix $\BY_K$ solves
\begin{align}
\label{eq:min}
\BY_K=\argmin_{\BZ\,\in\,\RNT}~~\|\BY - \BZ\|
~~\textrm{subj. to}~~\rank(\BZ)\le K \nonumber
\end{align}
where $\|\cdot\|$ can be the spectral or the Frobenius matrix norm. 
This minimizer attains the approximation errors
\begin{equation}\label{eq:errors}
\|\BY - \BY_K\|_2=\sigma_{K+1}~~\text{and}~~\|\BY - \BY_K\|_{\frob}^2=\sum_{k=K+1}^R\sigma_k^2.
\end{equation}
Evidently, $\BY_K$ approximates $\BY$ well if the $R-K$ trailing singular values are sufficiently small.
In practice, the rank $K$ is selected to deliver a relative approximation error below a certain threshold $0<\alpha<1$; for example, pick $K$ so that
\begin{align}%
\label{eq:chooseK}
    \frac{\|\BY-\BY_K\|_{\frob}}{\|\BY\|_{\frob}}\;=\;\frac{\big(\sum_{k=K+1}^R \sigma_k^2\big)^{1/2}}{\big(\sum_{k=1}^R \sigma_k^2\big)^{1/2}}\; \leq \; \alpha.
\end{align}
This compression via the SVD is akin to keeping the $K$ principal components of a matrix, as practiced in~\cite{XieCK14}.%
\footnote{Strictly speaking, the data would first be prepared for PCA by subtracting the mean of each row from every entry in that row, replacing $\BY$ with $\BY-\Bmu\Bone_{T}^{\trans}$, where $\mu_j = (y_{j,1} + \cdots + y_{j,T})/T$. We do not apply any such preprocessing of $\BY$, and thus take PCA to be synonymous with the SVD.}

As noted in the introduction,  the SVD blends information from \emph{all} PMU datastreams at \emph{all} times.
Because PMUs have limited capacity for handling data, measurements from every datastream must first be transmitted across dedicated synchrophasor communication links before the SVD can be applied to reduce the dimension of the data.
Thus, the SVD is not typically feasible for time-sensitive applications involving large-scale systems and is better suited for offline tasks, such as post-event analysis.

%%%%%%%%%%%%%%%%%%%%%%%%%%%%%%%%%%%%%%%%%%%%%%%%%%%%%%%%%%%%%%%%%%%%%%%%%%%%%%%%
\subsection{Interpolatory Approximations of PMU Data}
\label{ss:interpPMU}
\begin{figure}
\vspace*{14pt}
\includegraphics[height=0.55in]{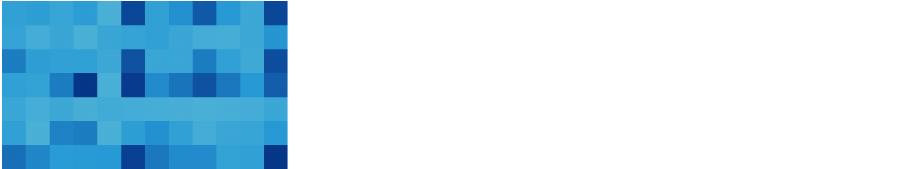}

\begin{picture}(0,0)
\put(30,58)   {\small $\BY$}
\put(113,57)  {\small $\BY\hphantom{_K}$ = original data matrix}
\put(113,44)  {\small $\BY_K$ = PCA / SVD approximation}
\put(113,31)  {\small $\BYs$\, = interpolatory approximation}
\put(138.5,20){\small for rows $\CS = \{4,7,6\}$}
\put(113,7)  {\small $\BYt$ = interpolatory approximation}
\put(138.5,-4) {\small for columns $\CT = \{12,4,9\}$}

\put(82,51)  {\rotatebox{-90}{\small \emph{streams $\to$}}}
\put(71,47)  {\scriptsize \emph{1}}
\put(71,40.5){\scriptsize \emph{2}}
\put(72,25)  {\scriptsize $\vdots$}
\put(71,13)  {\scriptsize \emph{7}}
\put(28,-4.5){\small \emph{time $\to$}}
\put(1,4.5)  {\scriptsize \emph{1}}
\put(7,4.5)  {\scriptsize \emph{2}}
\put(13,4.5) {\scriptsize \emph{3}}
\put(19,4.5) {\scriptsize \emph{4}}
\put(36,4.5) {\scriptsize $\cdots$}
\put(61,4.5) {\scriptsize \emph{1\!2}}

\end{picture}

\vspace*{30pt}
\includegraphics[height=0.55in]{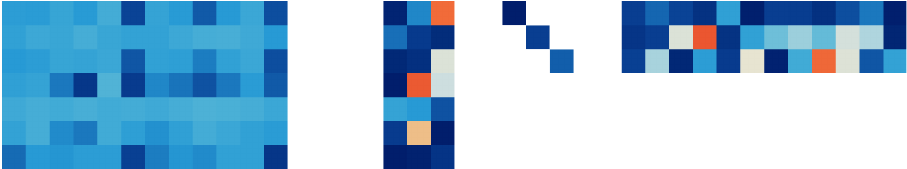}

\begin{picture}(0,0)
\put(30,58) {\small $\BY_K$}
\put(75,30) {\small $=$}
\put(91,58) {\small $\BU_K$}
\put(120,58){\small $\BSigma_K$}
\put(175,58){\small $\BV_K^{@\tr}$}
\put(120,23)  {\small \emph{$\BY_K$ approximates $\BY$ using}}
\put(120,13)  {\small \emph{the SVD (all streams, all times)}}
\end{picture}

\vspace*{15pt}
\includegraphics[height=0.55in]{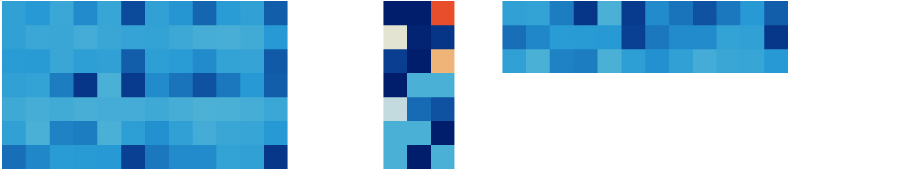}

\begin{picture}(0,0)
\put(30,58)   {\small $\BYs$}
\put(75,30)   {\small $=$}
\put(92,58)   {\small $\BZs$}
\put(146,58)  {\small $\BRs$}
\put(189,47)  {\scriptsize \emph{4}}
\put(189,40.5){\scriptsize \emph{7}}
\put(189,34)  {\scriptsize \emph{6}}
\put(120,23)  {\small \emph{$\BYs$ approximates $\BY$ using data}}
\put(120,13)  {\small \emph{from a few streams at all times.}}
\end{picture}

\vspace*{20pt}
\includegraphics[height=0.55in]{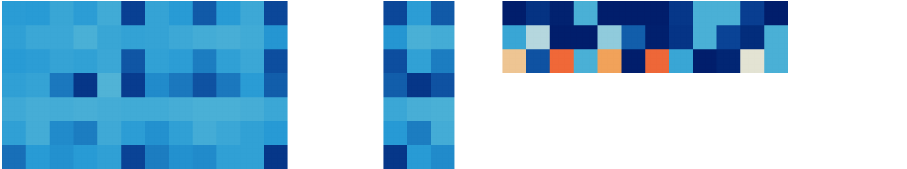}

\begin{picture}(0,0)
\put(30,58){\small $\BYt$}
\put(75,30){\small $=$}

\put(92,58) {\small $\BCt$}
\put(146,58){\small $\BWt$}

\put(88.7,4.5) {\scriptsize \textsf{1\kern-0.4pt 2}}
\put(97,4.5) {\scriptsize \textsf{4}}
\put(103,4.5){\scriptsize \textsf{9}}
\put(120,23) {\small \emph{$\BYt$ approximates $\BY$ using data}}
\put(120,13) {\small \emph{from all streams at a few times.}}
\end{picture}

\vspace*{20pt}
\includegraphics[height=0.55in]{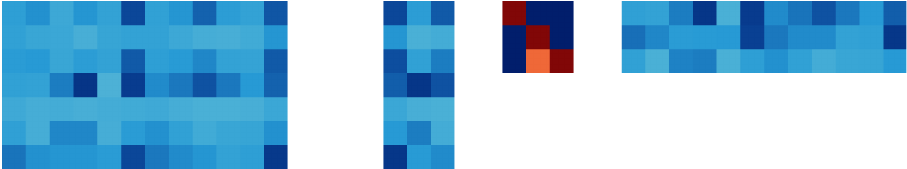}

\begin{picture}(0,0)
\put(30,58) {\small $\BYst$}
\put(75,30) {\small $=$}
\put(92,58) {\small $\BCt$}
\put(120,58){\small $\BXst$}
\put(175,58){\small $\BRs$}

\put(88.7,4.5)  {\scriptsize \textsf{1\kern-0.4pt 2}}
\put(97,4.5)  {\scriptsize \textsf{4}}
\put(103,4.5) {\scriptsize \textsf{9}}
\put(216,47)  {\scriptsize \emph{4}}
\put(216,40.5){\scriptsize \emph{7}}
\put(216,34)  {\scriptsize \emph{6}}
\put(120,23)  {\small \emph{$\BYst$ approximates $\BY$ using data}}
\put(120,13)  {\small \emph{from all streams at a few times,}}
\put(120,4)   {\small \emph{and a few streams at all times.}}
\end{picture}

\caption{\label{fig:XR}
Sketch of four low-rank approximations ($\BYk$, $\BYs$, $\BYt$, and $\BYst$) to the PMU data matrix $\BY$.\ \ The bottom three are IDs.  (See Figure~2 of~\cite{Hen24} for a similar graphic illustrating the DEIM process.)
}
\end{figure}

%%%%%%%%%%%%%%%%%%%%%%%%%%%%%%%%%%%%%%%%%%%%%%%%%%%%%%%%%%%%%%%%%%%%%%%%%%%%%%%%
As an alternative to PCA/SVD for PMU data, we propose using \emph{interpolatory matrix decompositions} (IDs)~\cite{MahD09, SorE16, LWMRT07, DonM23, Ste99}.
These low-rank factorizations, sketched in Figure~\ref{fig:XR}, use only a few rows and/or columns of the matrix.
For $K\leq N$, an ID of $\BY\in\R^{N\times T}$ is a low-rank factorization of the form
\begin{equation}
\label{eq:interpApprox}
    \BY\approx\BYst\coloneqq\BCt@@ \BXst@@\BRs^{}
\end{equation}
where $\BXst\in\R^{K\times K}$. The matrices $\BCt\coloneqq \BY(:,\,\CT) \in\R^{N\times K}$ and $\BRs\coloneqq\BY(\CS,\,:)\in\R^{K\times T}$
contain a subset of the columns and rows of $\BY$ indexed by $\CT=\{t_1, t_2,\ldots,t_K\}\subset\{1,2,\ldots, T\}$ and $\CS=\{s_1, s_2,\ldots, s_K\}\subset\{1,2,\ldots,N\}$. We refer to the PMU datastreams and time snapshots corresponding to the indices $\CS$ and $\CT$ as \emph{pilot streams} and \emph{pilot snapshots}, respectively, or simply \emph{pilots} when referring to both. The ID in~\eqref{eq:interpApprox} aims to recover the complete matrix $\BY$ using only information contained in the selected columns $\BCt$ and rows $\BRs$ of $\BY$; the matrix $\BXst$ of smaller dimension $K \times K$ is chosen to make $\BYst$ a good approximation of $\BY$.

The approximation quality hinges on which rows and columns are selected for $\BRs$ and $\BCt$, and the choice for $\BXst$. In Section~\ref{ss:interpError}, we describe a strategy for computing $\BXst$ via a least-squares fit of the data, once $\CS$ and $\CT$ are set. The numerical linear algebra literature has explored various strategies for selecting rows $\CS$ and columns $\CT$; see, e.g.,~\cite{MahD09, SorE16, DonM23, LWMRT07}.
In Section~\ref{sec:DEIM}, we propose a greedy strategy from~\cite{ChaS10, Baretal04} to select rows and columns iteratively.

%%%%%%%%%%%%%%%%%%%%%%%%%%%%%%%%%%%%%%%%%%%%%%%%%%%%%%%%%%%%%%%%%
%%%%%%%%%%%%%%%%%%%%%%%%%%%%%%%%%%%%%%%%%%%%%%%%%%%%%%%%%%%%%%%%%

The form~\eqref{eq:interpApprox} is the most general; one may also consider IDs that only use the rows \emph{or} columns of $\BY$.\ \ Utilizing only certain rows of $\BY$ amounts to using data collected from $K\leq N$ pilot datastreams contained in $\BRs$ to approximate the data from all datastreams, i.e.,
finding a matrix $\BZs\in\R^{N\times K}$ such that 
\begin{align}
    \label{eq:interpRowApprox}
    \BY \approx \BYs \coloneqq\BZs\BRs \in \R^{N\times T}.
\end{align}
Note that in~\eqref{eq:interpRowApprox}, columns of $\BY$ are not factored into the approximation. The $i$-th row of $\BZs$ contains the weights that specify how the data collected from the $K$ pilot streams in $\BRs$ should be combined to approximate the data $\BY(i,\,:)$ from the $i$-th \emph{non-pilot} datastream, $i\not\in\CS$:
\begin{equation}
\label{eq:interpRowApprox_oneRow}
    \BY(i,\,:\,)\approx \left(\BZs\BRs\right)(i,\,:\,) = \sum_{k=1}^K \BZs(i,k) \BRs(k,\,:\,).
\end{equation}

Using only certain columns of $\BY$ amounts to using the $K\leq T$ pilot snapshots contained in $\BCt$ to recover the full time series, i.e., finding a matrix $\BWt\in\R^{K\times T}$ such that 
\begin{equation}
    \label{eq:interpColApprox}
    \BY \approx \BYt \coloneqq \BCt\BWt \in \R^{N\times T}\kern-2pt.
\end{equation}
Now the $j$-th column of $\BWt$ contains the weights that describe how the selected pilot snapshots in $\BCt$ should be combined to produce an approximation of the data $\BY(:\,,j)$ at time $t_j$ for $j\not\in\CT$, i.e.,
\begin{equation}
\label{eq:interpColApprox_oneCol}
    \BY(\,:\,,j)\approx \left(\BCt\BWt\right)(\,:\,,j) = \sum_{k=1}^K \BWt(j,k)\BCt(\,:\,,k).
\end{equation}

Figure~\ref{fig:XR} compares
different ID regimes to the SVD.\ \ 
The storage requirements for $\BCt$, $\BXst$, and $\BRs$ are similar to that of the SVD.\ \ 
Like the optimal approximation $\BY_K$ from the SVD, the interpolatory approximations $\BYst$, $\BYs^{}$, and $\BYt$ have rank $K$ (or less).
Unlike the orthogonal components $\BU_K$ and $\BV_K$ in the SVD, the low-rank factors $\BRs$ and $\BCt$ contain actual PMU data.
These factors preserve qualitative features of the data; e.g., sparsity, or a particular voltage pattern if the data was collected following a disturbance~\cite{KonFYY21}.

Another virtue of the interpolatory approximations over $\BY_K$ is economy: $\BY_K$ is a blend of 
\emph{all} $N\gg K$ rows and $T\gg K$ columns of $\BY$, so computing the SVD requires information from \emph{all} PMU time series simultaneously.
On the other hand, interpolatory approximations $\BYs$ and $\BYt$ only use $K=|\CS|=|\CT|$ rows and columns of $\BY$.\ \  
Once the weights $\BZs$ and $\BWt$ have been computed, a low-rank approximation to $\BY$ can be obtained while interacting \emph{only} with $K\leq N$ pilot streams or $K\leq T$ down-sampled pilot snapshots. As observed in, e.g.,~\cite{XieCK14, DasS13, Das16}, this reduction significantly lowers the bandwidth needs of synchrophasor communication networks. 
We point out that the sparse approximations of PMU data proposed in~\cite{XieCK14, Liuetal19, DasS13, Das16} can be interpreted at the matrix level as one-sided interpolatory approximations~\eqref{eq:interpRowApprox} and~\eqref{eq:interpColApprox}.

%%%%%%%%%%%%%%%%%%%%%%%%%%%%%%%%%%%%%%%%%%%%%%%%%%%%%%%%%%%%%%%%%
\subsection{Analyzing the Interpolatory Approximation Error}
\label{ss:interpError}
%%%%%%%%%%%%%%%%%%%%%%%%%%%%%%%%%%%%%%%%%%%%%%%%%%%%%%%%%%%%%%%%%

Because $\BYk$ is the optimal rank-$K$ approximation to $\BY$,
the interpolatory approximation $\BYst$ cannot be any better:
\begin{equation} \label{eq:interplb}
    \sigma_{K+1}  = \|\BY - \BY_K\|_2 \leq \|\BY-\BYst\|_2.
\end{equation}
The same holds for  $\BYs$ and $\BYt$.
We analyze here the quality of the approximation $\BYst$ relative to the best approximation from PCA/SVD.
For the moment, we assume the $K$ row and column indices in $\CS$ and $\CT$ are given.
The error $\|\BY-\BYst\|_2$ depends on how we compute $\BXst$.\ \ One natural choice~\cite{MahD09,Ste99} is
\begin{equation*}
    \BXst = \left(\BCt\right)^{\dagger}\BY\big(\BRs\big)^{\dagger}
\end{equation*}
where $\cdot^\dagger$ denotes the \emph{Moore--Penrose pseudoinverse}; see, e.g., Section 5.5.2 of~\cite{GolV12}. Assuming the rows of $\BRs$ and the columns of $\BCt$ are linearly independent, we have $\left(\BRs\right)^{\dagger} = \BRs^{\trans} \left(\BRs^{}\BRs^{\trans}\right)^{-1}$ and $\left(\BCt\right)^\dagger =\big({\BCt}^{\trans}{\BCt}\big)^{-1}{\BCt}^{\trans}.$
Thus, $\BYst$ is given by
\begin{equation*}
    \BYst = \BCt\BXst\BRs = \BCt\left(\BCt\right)^{\dagger}\BY\left(\BRs\right)^\dagger \BRs.
\end{equation*}
For the one-sided interpolatory approximations $\BYs$ and $\BYt$, the same idea can be applied to obtain $\BZs$ and $\BWt$:
\begin{equation*}
\BZs=\BY\left(\BRs\right)^\dagger~~\mbox{and}~~\BWt=\left(\BCt\right)^{\dagger}\BY.
\end{equation*}
This construction highlights why IDs can effectively  compress the entire PMU data set in $\BY$: the matrices $\BZs$ and $\BWt$ are least squares fits of the data~\cite{Ste99}, e.g.,
\begin{align}
\begin{split}
    \label{eq:lsSol}
    \BZs&\,=\, \argmin_{\BZ\in\R^{N\times K}}\|\BY-\BZ\BRs\|_{\frob}.
\end{split}
\end{align}
Assuming the rows (columns) of $\BY$ are linearly independent, the solutions $\BZs$ ($\BWt$) to~\eqref{eq:lsSol} are unique.
The pilot-based reconstruction from~\cite{XieCK14} in fact uses this choice of $\BZs$.

The quality of the IDs $\BYst$, $\BYs$, and $\BYt$ can be assessed in a more quantitative way. Define $\BS \coloneqq \BI_N(\,:\,,\CS)\in\R^{N\times K}$ and $\BT\coloneqq\BI_T(\,:\,,\CT)\in\R^{T\times K}$ to be the matrices containing the $K$ columns of the $N\times N$ and $T \times T$ identity matrices indexed by $\CS$ and $\CT$.\ \ 
From Theorem 4.1 of~\cite{SorE16}, we have that
\begin{equation} 
    \label{eq:interpErrorBound}
    \sigma_{K+1} \ \leq\  \|\BY-\BYst\|_2\ \leq\ (\etaBs+\etaBt) \;\sigma_{K+1}
\end{equation}
where error factors $\etaBs,\etaBt\geq 1$ are given by
\begin{equation}
\label{eq:errorConstants}
    \etaBs \coloneqq \|\left(\BS^{\trans}\BU_K\right)^{-1}\|_2 ~~\mbox{and}~~
    \etaBt \coloneqq \|\left(\BT^{\trans}\BV_K\right)^{-1}\|_2.
\end{equation}
Recall that $\BU_K\in\R^{N\times K}$ and $\BV_K\in\R^{T\times K}$ contain the leading $K$ left and right singular vectors of $\BY$.\ \ 
The submatrices $\BS^{\trans}\BU_K$ and $\BT^{\trans}\BV_K$ are guaranteed to be nonsingular for certain row and column selection schemes, including the one we propose in Section~\ref{sec:DEIM}; see Lemma~3.2 of~\cite{SorE16}.
For the one-sided interpolatory approximations $\BYs$ and $\BYt$, we have the simplified bounds:
\begin{align}
\begin{split}
\label{eq:oneSidedInterpErrorBound}
    \sigma_{K+1}\  &\leq\   \|\BY-\BYs\|_2\ \leq\ \etaBs\, \sigma_{K+1}\\
    \sigma_{K+1}  \ &\leq\   \|\BY-\BYt\|_2\ \leq\ \etaBt\,\sigma_{K+1}.
\end{split}
\end{align}
For details, see Lemma~4.2 of \cite{SorE16} and Theorem~1.5 of~\cite{HonP92}.

Let us unpack the error factors $\etaBs$ and $\etaBt$.\ \ 
The matrix $\BS^{\trans}\BU_K$ is a $K\times K$ submatrix of $\BU_K$.\ \ Note that $\BU_k$ has orthonormal columns;
$\etaBs$ measures how far from orthonormal the
\emph{rows} of $\BU_K$ corresponding to $\CS$ are.  
Likewise, $\etaBt$ measures how far from orthonormal the $\CT$ rows of $\BV_K$ are.
The order of the indices
in $\CS$ and $\CT$ does not affect $\etaBs$ and $\etaBt$.

The interpolatory error bounds in~\eqref{eq:interpErrorBound} and~\eqref{eq:oneSidedInterpErrorBound} hold for \emph{any} collection of pilot indices $\CS$ or $\CT$. We can leverage these bounds to realize real-time computational benefits and performance guarantees. Although these ideas also apply to the two-sided formulation in~\eqref{eq:interpApprox}, we present them for the one-sided approximations~\eqref{eq:interpRowApprox} and~\eqref{eq:interpColApprox} for simplicity.
\begin{enumerate}
    \item[{1.}] \emph{Fast error monitoring}. 
    Because $\BS^{\trans} \BU_K$ and $\BT^{\trans}\BV_K$ are small $K\times K$ matrices, the error
    indicators $\etaBs$ and $\etaBt$ will be much quicker to compute than the
    full approximation errors $\|\BY-\BYs\|_2$ or $\|\BY-\BYt\|_2$ for large $N$ and $T$.\ \ 
    One does not even need to explicitly form the low-rank approximations $\BYs$ or $\BYt$ in~\eqref{eq:interpRowApprox} and~\eqref{eq:interpColApprox}, and hence $\BZs$ or $\BWt$, to evaluate $\etaBs$ and $\etaBt$.\ \ 
    This observation allows for fast \emph{a priori} estimation of the interpolatory approximation error during online operations, or enables the error factors to be monitored as the pilots are selected.
    \item[{2.}] \emph{Pilot certification.} 
    In an operational setting, one can use any desired strategy for
    picking pilots $\CS$ and $\CT$.\ \ The error factors $\etaBs$ or $\etaBt$ can then be (quickly) computed to certify if the chosen pilots capture the rank-$K$ nature of the PMU data matrix. If the error bound is below a threshold, e.g.,
    $\etaBs\sigma_{K+1},\,\etaBt\sigma_{K+1} \leq \tau = 10^{-1}$, the selection $\CS$ or $\CT$ is accepted; otherwise, either replace some pilots, or increase $K$ and add additional pilots. 
\end{enumerate}
We describe at length how each of the above ideas can be implemented in a practical operational scenario in Section~\ref{sec:eventMonitoring}.

One could consider selecting pilots $\CS$ and $\CT$ to explicitly minimize $\etaBs$ and $\etaBt$ over all possible configurations; however, such a minimization would involve combinatorial complexity.
Instead, in Section~\ref{sec:DEIM} we
advocate for a more efficient \emph{greedy} algorithm that seeks to
control the growth of the error factors $\etaBs$ and $\etaBt$ as new pilots are selected, one at a time.

% SECTION 3
%%%%%%%%%%%%%%%%%%%%%%%%%%%%%%%%%%%%%%%%%%%%%%%%%%%%%%%%%%%%%%%%%%%
\section{Greedy Pilot Selection}
\label{sec:DEIM}
%%%%%%%%%%%%%%%%%%%%%%%%%%%%%%%%%%%%%%%%%%%%%%%%%%%%%%%%%%%%%%%%%%%
We propose using the \emph{discrete empirical interpolation method} (DEIM) index selection algorithm~\cite{Baretal04,ChaS10,SorE16,DrmG16,Hen24} to select the pilot subsets $\CS$ and $\CT$.\ \ DEIM is a discrete variant of the empirical interpolation method~\cite{Baretal04}. Initially developed for resolving the ``lifting bottleneck'' in the model reduction of nonlinear dynamical systems~\cite{ChaS10} by constructing interpolatory approximations to vector-valued nonlinear functions, DEIM was applied to construct IDs in~\cite{SorE16}. The DEIM procedure (independently) selects the row and column indices $\CS$ and $\CT$ by iteratively parsing the leading left and right singular vectors stored in $\BU_K$ and $\BV_K$ for a matrix $\BY$.\ \ At each iteration, DEIM attempts to adaptively minimize the growth of the error factors in~\eqref{eq:errorConstants} as each new index is added to $\CS$ or $\CT$, and, in practice, the DEIM indices typically yield small error factors. In the numerical tests of Section~\ref{sec:eventMonitoring}, DEIM selects pilot configurations that produce error factors $\etaBs$ and $\etaBt$ of size $\mathcal{O}(10^1)$ or less, whereas other, seemingly reliable, selection approaches produce factors of size $\mathcal{O}(10^4)$.
Thus, in conjunction with the approximation error~\eqref{eq:interpErrorBound}, we expect DEIM to provide an effective pilot selection strategy.

%%%%%%%%%%%%%%%%%%%%%%%%%%%%%%%%%%%%%%%%%%%%%%%%%%%%%%%%%%%%%%%%%%%
\subsection{The Discrete Empirical Interpolation Method}
%%%%%%%%%%%%%%%%%%%%%%%%%%%%%%%%%%%%%%%%%%%%%%%%%%%%%%%%%%%%%%%%%%%
We describe how DEIM operates on $\BU_K$ to select $K\leq N$ pilot streams $\CS$ from a data matrix $\BY$.\ \ The same process applied (independently) to $\BV_K$ selects the pilot snapshots $\CT$.

Our derivation uses special matrices called \emph{interpolatory projectors}. 
Let $\BS_k=\begin{bmatrix}
    \Be_{s_1} & \cdots & \Be_{s_k}
\end{bmatrix}\in\R^{N\times k}$ denote the $k$ columns of $\BI_N$ specified by the distinct indices in $\CS_k=\{s_1,\ldots,s_k\}\subset\{1,\ldots,N\}$, and let $\BU_k=\BU(@:@, {1@:@k})= \begin{bmatrix}\Bu_1 & \cdots & \Bu_k\end{bmatrix} \in\R^{N\times k}$ denote the leading $k$ columns of the matrix $\BU\in\R^{N\times R}$ of the left singular vectors of $\BY$.\ \ 
The \emph{interpolatory projector} for $\CS_k$ onto $\mspan\left(\BU_k\right)$ is defined as
\begin{equation}
    \label{eq:interpProj}
    \BP_k\coloneqq \BU_k\left(\BS_k^{\trans}\BU_k^{}\right)^{-1}\BS_k^{\trans}\in\R^{N\times N}\kern-2pt.
\end{equation}
The matrix $\BS_k^{\trans}\BU_k^{}\in\R^{k\times k}$ is guaranteed to be invertible for indices $\CS_k$ adaptively selected by DEIM (see Lemma~3.2 of~\cite{SorE16}). 
One can readily verify that $\BP_k$ satisfies the projector property: $\BP_k^2=\BP_k$.\ \ 
More critically,  $\BP_k$ is an \emph{interpolatory} projector in this sense: 
for any vector $\Bx\in\R^N$, the projected vector $\Bxhat\coloneqq\BP_k\Bx$ exactly matches $\Bx$  in the pilot indices $\CS_k$, i.e., 
\begin{equation}
    \label{eq:interpProp}
\Bxhat(\CS_k)=\BS_k^{\trans} \left(\BU_k^{} (\BS_k^{\trans}\BU_k^{})^{-1}\BS_k^{\trans}\right)\Bx = \Bx(\CS_k).
\end{equation}

DEIM operates on the columns of $\Bu_k$ one at a time to select each new pilot index. Start with the selection of the first pilot $s_1$, corresponding to $k=1$. In this simple case, the error factor $\eta_{\CS_1}$ in~\eqref{eq:errorConstants} reduces to 
\begin{equation*}
    \eta_{\CS_1} = \Bigl\|\left(\BS_1^{\trans}\BU_1^{}\right)^{-1}\Bigr\|_2=\frac{1}{|\Bu_1(s_1)|}
\end{equation*}
the reciprocal of the magnitude of the $s_1$ entry 
of the leading singular vector $\Bu_1$. Therefore, to minimize $\eta_{s_1}$, choose the datastream index $s_1\in\{1,\ldots, N\}$ corresponding to the entry in $\Bu_1 \in\R^N$ having the largest magnitude. 

\smallskip
\begin{itemize}
    \item \textbf{Step~1.} Choose $s_1$ as the index corresponding to the entry of $\Bu_1$ with the largest magnitude:
    \begin{equation*}
        s_1 = \argmax_{1\leq s \leq N}|\Bu_1(s)|, \quad 
        \Bu_1 = \left[\begin{array}{c}
        \times\\
        \hot{\times}\\
        \times\\
        \times\\
        \end{array}\right]
        \begin{array}{c}
        \\
        \!\!\!\leftarrow s_1\\
        \\
        \\
        \end{array}.
    \end{equation*}
    Construct $\BP_1 := \Bu_1^{} \Be_{s_1}^{\trans} / \Bu_1(s_1)$, the interpolatory projector~\eqref{eq:interpProj}  
    for $\CS_1=\{s_1\}$ onto the span of $\Bu_1$.  
\end{itemize}

The choice of the second index $s_2$ is more subtle. We should avoid choosing the same pilot ($s_2=s_1$), which would result in an infinite error factor $\etaBs=\|\left(\BS_2^{\trans}\BU_2\right)^{-1}\|_2$.
Using the intuition that $\etaBs=\|\left(\BS_k^{\trans}\BU_k\right)^{-1}\|$ is small when the rows of $\BU_k$ selected by $\BS_k$ are quite distinct, we choose $s_2$ so that the two rows $\BU_2(\CS_2,\,:\,)$ are as \emph{independent as possible} for $\CS_2=\{s_1,s_2\}.$
To guarantee that $s_2 \neq s_1$, i.e., that we select a distinct datastream, we remove a multiple of $\Bu_1$ from $\Bu_2$ to zero out the $s_1$ entry:
\begin{equation*}
    \Br_2 \coloneqq \Bu_2 - \frac{\Bu_2(s_1)}{\Bu_1(s_1)} \Bu_1 = \Bu_2 - \BP_1 \Bu_2
\end{equation*}
giving $\Br_2(s_1) = 0$ by the interpolatory property of $\BP_1$ in index $s_1$. We can then select $s_2$ to be the index of the largest-magnitude entry of $\Br_2$.
Using this formulation, we summarize the next step of DEIM as follows.

\smallskip
\begin{itemize}
    \item \textbf{Step 2.}
    Compute the residual of the interpolatory 
    projection of $\Bu_2$ onto ${\rm span}\left(\Bu_1\right)$:
    \begin{equation*}
        \Br_2 = \Bu_2 -\BP_1 \Bu_2.
    \end{equation*}
    Choose $s_2$ as the largest-magnitude entry of $\Br_2$:
    \begin{equation*}
        s_2 = \argmax_{1\leq n \leq N}|\Br_2(n)|, \quad
        \Br_2 = \Bu_2 - \BP_1\Bu_2 =\left[\begin{array}{c}
        \star\\
        0\\
        \star\\
        \hot{\star}\\
        \end{array}\right]
        \begin{array}{c}
        \\
        \\
        \\
        \!\!\!\leftarrow s_2\\
        \end{array}
    \end{equation*}
    (The $\star$ indicates a modified entry from the $k=1$ step.)
\end{itemize}
\smallskip
Subsequent steps, $k= 3,\ldots, K$, follow this same template.
\begin{itemize}    
    \item \textbf{Step~\boldmath $k$.} Construct the interpolatory 
    projector $\BP_{k-1}$ for datastreams $s_1,\ldots, s_{k-1}$ onto the span of 
    $\Bu_1, \ldots, \Bu_{k-1}$ according to~\eqref{eq:interpProj}, and
    compute the residual
    \begin{equation*}
        \Br_k = \Bu_k -\BP_{k-1} \Bu_k
    \end{equation*}
    such that $\Br_k(s_1) = \cdots = \Br_k(s_{k-1}) = 0$.
    Choose $s_k$ to be the index of the largest-magnitude entry of $\Br_k$:
    \begin{equation*}
        s_k = \argmax_{1\leq s \leq N}|\Br_k(s)|.
    \end{equation*}
\end{itemize}
We are assured that 
$s_k$ is a new datastream that differs from $s_1,\ldots, s_{k-1}$, 
since $\Br_k(s_1) = \cdots = \Br_k(s_{k-1}) = 0$ 
but $\Br_k \neq \Bzero$
(otherwise, $\Bu_k \in {\rm span}\{\Bu_1, \ldots, \Bu_{k-1}\}$, a contradiction).

%%%%%%%%%%%%%%%%%%%%%%%%%%%%%%%%%%%%%%%%%%%%%%%%%%%%%%%%%%%%
\begin{algorithm2e}[t]
  \SetAlgoHangIndent{1pt}
  \DontPrintSemicolon
  \caption{The discrete empirical interpolation method (DEIM)~\cite{ChaS10,Baretal04}.} 
  \label{alg:DEIM}
  \KwIn{Matrix with orthonormal columns $\BU_K=\begin{bmatrix}
      \Bu_1 & \cdots & \Bu_K
  \end{bmatrix}\in\R^{N\times K}$, $1\leq K<N$.}
  \KwOut{Indices $\CS=\{s_1,\ldots,s_K\}\subset\{1,\ldots,N\}$.}
  Choose the first index $s_1 = \argmax_{1\leq s \leq N}|\Bu_1(s)|.$\;
  Take $\CS_1= \{s_1\}$.\;
  \For{$k = 2, \ldots,K$}{
    Compute the residual by solving a $k$-dimensional linear system:
    \begin{equation*}
        \Br_k=\Bu_k - \BU_{k-1}\left(\BS_{k-1}^{\trans}\BU_{k-1}\right)^{-1}\BS_{k-1}^{\trans}\Bu_{k}.
    \end{equation*}
    \;
    \vspace{-4mm}
    Choose $s_k = \argmax_{1\leq s \leq N}|\Br_k(s)|$.\;
  Take $\CS_k= \CS_{k-1} \cup \{s_k\}$.
  }
\end{algorithm2e}
%%%%%%%%%%%%%%%%%%%%%%%%%%%%%%%%%%%%%%%%%%%%%%%%%%%%%%%%%%%%

Algorithm~\ref{alg:DEIM} summarizes this procedure.
(To select pilot snapshots $\CT$, i.e., columns of $\BY$, simply apply this algorithm to the right singular vectors $\BV_K$.)\ \ 
An efficient implementation avoids explicitly forming the interpolatory projectors $\BP_k$, which are large, dense matrices. Rather, step~4 of Algorithm~\ref{alg:DEIM} computes the \emph{action} of $\BP_k$ on the singular vector $\Bu_k$ without explicitly forming the inverse of $\BS_{k-1}^{\trans}\BU_{k-1}$. Instead, a $(k-1)\times(k-1)$ linear system of equations is solved at every step. These systems can be solved efficiently by leveraging the nested structure of the coefficient matrix $\BS_{k-1}^{\trans}\BU_{k-1}$ to compute its LU decomposition.
In such an implementation, Algorithm~\ref{alg:DEIM} can be carried out in $\CO(NK^2) + \CO(K^3)$ floating point operations (FLOPs). 
We refer to Section 2.1.2 of~\cite{DrmG16} for a more detailed complexity analysis.
Typically, $K \ll N$ in practice. Given the requisite singular value and vector data (also required for PCA-based methods), the cost of DEIM is linear in $N$ (or analogously $T$, when used to select column indices). Thus, DEIM can be realistically applied for pilot selection in large-scale settings.

At every iteration, the DEIM selection is designed to roughly minimize the incremental growth of the error factor $\etaBs$ in the bound~\eqref{eq:oneSidedInterpErrorBound}; see Lemma~3.2 in~\cite{ChaS10} for a proof. 
This explains why DEIM is an effective choice for computing the pilot sets $\CS$ and $\CT$, as illustrated in Section~6 of~\cite{SorE16}.
The DEIM algorithm is directly linked to the LU factorization with partial pivoting; see Section~3 of~\cite{SorE16}.
One alternative to the DEIM index selection algorithm is the QDEIM variant~\cite{DrmG16}, which
identifies the pilots $\CS$ by applying a
rank-revealing QR factorization to the rows of $\BU_K$.\ \ 
The cost of the factorization is $2NK^2 - \tfrac{2}{3}K^3$ FLOPs (to leading order); see, e.g., Section~5.4.3 of~\cite{GolV12}. Thus, the cost of QDEIM is similar to that of DEIM.
The ultimate set of pilots $\CS$ chosen by QDEIM is invariant under permutations of the columns of $\BU_K$, although in practice, DEIM and QDEIM perform similarly.
We emphasize that QDEIM is not iterative; the number of desired pilots must be specified in advance.

%%%%%%%%%%%%%%%%%%%%%%%%%%%%%%%%%%%%%%%%%%%%%%%%%%%%%%%%%%%%%%%%%%%
\subsection{Numerical Tests}
\label{ss:interpNumerics}
%%%%%%%%%%%%%%%%%%%%%%%%%%%%%%%%%%%%%%%%%%%%%%%%%%%%%%%%%%%%%%%%%%%
%%%%%%%%%%%%%%%%%%%%%%%%%%%%%%%%%%%%%%%%%%%%%%%%%%%%%%%%%%%%%%%%%
\begin{figure}[t!]
\centering
  \tikzexternalenable%
  \tikzsetnextfilename{interpApproxLegend}%
  \begin{tikzpicture}
  \begin{axis}[%
    hide axis,
    width  = 1mm,
    height = 1mm,
    scale only axis,
    xmin = 0,
    xmax = 1,
    ymin = 0,
    ymax = 1,
    legend columns = 5, 
    legend style   = {
      at     = {(0,0)},
      anchor = center,
      /tikz/every even column/.append style = {column sep = 0.1cm}},
    legend cell align  = {left},
    clip mode          = individual,
    cycle list name    = interpApproxPlotlist]

    \foreach \y in {1, 2, ..., 5}{
      \addplot+ coordinates{ (0, 0) };
    }
    \plotfontsize
    \addlegendentry{$\sigma_{k+1}/\sigma_{1}$}
    \addlegendentry{DEIM}
    \addlegendentry{QDEIM}
    \addlegendentry{MILP}
    \addlegendentry{RAND}
    % \addlegendentry{...}
  \end{axis}
\end{tikzpicture}%
  \tikzexternaldisable%

\begin{subfigure}[b!]{\linewidth}
    \raggedright
  \tikzexternalenable%
  \tikzsetnextfilename{interpRowApprox_voltage}%
  \begin{tikzpicture}[font = \plotfontsize]
  \pgfplotstableread{graphics/data/rowID_errors_busV.dat}\tableINPUT
  
  \begin{semilogyaxis}[%
    width  = .75\linewidth,
    height = .1\textheight,
    scale only axis,
    grid=both,
    grid style={line width=.1pt, draw=gray!10},
    major grid style={line width=.2pt,draw=gray!25},
    minor tick num=4,
    xmin = 1,
    xmax = 20,
    xtick = {1, 5, 10, 15, 20},
    extra x ticks = {2, 3, 4, 6, 7, 8, 9, 11, 12, 13, 14, 16, 17, 18, 19},
    extra x tick labels={},
    ymin = 5*1e-6,
    ymax = 2*1e-1,
    xminorticks = true,
    yminorticks = true,
    xlabel = {rank, $K$},
    ylabel = {$\|\BYvoltage-\BYt\|_2/\|\BYvoltage\|_2$},
    ylabel style   = {yshift = -.3em},
    scaled x ticks = false,
    x tick label style = {/pgf/number format/1000 sep={\,}},
    y tick label style = {/pgf/number format/1000 sep={\,}},
    cycle list name    = interpApproxPlotlist
  ]
  
    \foreach \y in {1, ..., 5}{
      \addplot+ [restrict x to domain=0:20] table[x index = 0, y index = \y] {\tableINPUT};
    }

    \node[anchor=north east, font=\small] at (rel axis cs:1,.97) {Row-based $\BYs \approx \BYvoltage$ (voltage mag.)};
  \end{semilogyaxis}
\end{tikzpicture}%
  \tikzexternaldisable%
\\
    \vspace{.5\baselineskip}
  \tikzexternalenable%
  \tikzsetnextfilename{interpColApprox_voltage}%
  \begin{tikzpicture}[font = \plotfontsize]
  \pgfplotstableread{graphics/data/colID_errors_busV.dat}\tableINPUT
  
  \begin{semilogyaxis}[%
    width  = .75\linewidth,
    height = .1\textheight,
    scale only axis,
    grid=both,
    grid style={line width=.1pt, draw=gray!10},
    major grid style={line width=.2pt,draw=gray!25},
    minor tick num=4,
    xtick = {1, 5, 10, 15, 20},
    extra x ticks = {2, 3, 4, 6, 7, 8, 9, 11, 12, 13, 14, 16, 17, 18, 19},
    extra x tick labels={},
    xmin = 1,
    xmax = 20,
    ymin = 5*1e-6,
    ymax = 2*1e-1,
    xminorticks = true,
    yminorticks = true,
    xlabel = {rank, $K$},
    ylabel = {$\|\BYvoltage-\BYt\|_2/\|\BYvoltage\|_2$},
    ylabel style   = {yshift = -.3em},
    scaled x ticks = false,
    x tick label style = {/pgf/number format/1000 sep={\,}},
    y tick label style = {/pgf/number format/1000 sep={\,}},
    cycle list name    = interpColApproxPlotlist
  ]
  
    \foreach \y in {1, 2, 3, 4}{
      \addplot+ [restrict x to domain=0:20] table[x index = 0, y index = \y] {\tableINPUT};
    }

    \node[anchor=north east, font=\small] at (rel axis cs:1,.97) {Column-based $\BYt \approx \BYvoltage$ (voltage mag.)};
  \end{semilogyaxis}
\end{tikzpicture}%
  \tikzexternaldisable%

    \caption{Relative errors for $K=1,2,\ldots,20$ interpolatory matrix approximations $\BYs$ and $\BYt$ of the voltage magnitude data $\BYvoltage.$}
    \label{fig:interpApprox_voltage}
    \vspace{.5\baselineskip}
\end{subfigure}
\vspace{1.5\baselineskip}
\begin{subfigure}[b!]{\linewidth}
\raggedright
  \tikzexternalenable%
  \tikzsetnextfilename{interpRowApprox_angle}%
  \begin{tikzpicture}[font = \plotfontsize]
  \pgfplotstableread{graphics/data/rowID_errors_thetaV.dat}\tableINPUT
  
  \begin{semilogyaxis}[%
    width  = .75\linewidth,
    height = .1\textheight,
    scale only axis,
    grid=both,
    grid style={line width=.1pt, draw=gray!10},
    major grid style={line width=.2pt,draw=gray!25},
    minor tick num=4,
    xmin = 1,
    xmax = 20,
    xtick = {1, 5, 10, 15, 20},
    extra x ticks = {2, 3, 4, 6, 7, 8, 9, 11, 12, 13, 14, 16, 17, 18, 19},
    extra x tick labels={},
    ymin = 5*1e-6,
    ymax = 1e0,
    xminorticks = true,
    yminorticks = true,
    xlabel = {rank, $K$},
    ylabel = {$\|\BYangle-\BYs\|_2/\|\BYangle\|_2$},
    ylabel style   = {yshift = -.3em},
    scaled x ticks = false,
    x tick label style = {/pgf/number format/1000 sep={\,}},
    y tick label style = {/pgf/number format/1000 sep={\,}},
    cycle list name    = interpApproxPlotlist
  ]
  
    \foreach \y in {1, ..., 5}{
      \addplot+ [restrict x to domain=0:20] table[x index = 0, y index = \y] {\tableINPUT};
    }

    \node[anchor=north east, font=\small] at (rel axis cs:1,.97) {Row-based $\BYs \approx \BYangle$ (phasor angle)};
  \end{semilogyaxis}
\end{tikzpicture}%
  \tikzexternaldisable%
\\
    \vspace{.5\baselineskip}
  \tikzexternalenable%
  \tikzsetnextfilename{interpColApprox_angle}%
  \begin{tikzpicture}[font = \plotfontsize]
  \pgfplotstableread{graphics/data/colID_errors_thetaV.dat}\tableINPUT
  
  \begin{semilogyaxis}[%
    width  = .75\linewidth,
    height = .1\textheight,
    scale only axis,
    grid=both,
    grid style={line width=.1pt, draw=gray!10},
    major grid style={line width=.2pt,draw=gray!25},
    minor tick num=4,
    xtick = {1, 5, 10, 15, 20},
    extra x ticks = {2, 3, 4, 6, 7, 8, 9, 11, 12, 13, 14, 16, 17, 18, 19},
    extra x tick labels={},
    xmin = 1,
    xmax = 20,
    ymin = 5*1e-6,
    ymax = 1e0,
    xminorticks = true,
    yminorticks = true,
    xlabel = {rank, $K$},
    ylabel = {$\|\BYangle-\BYt\|_2/\|\BYangle\|_2$},
    ylabel style   = {yshift = -.3em},
    scaled x ticks = false,
    x tick label style = {/pgf/number format/1000 sep={\,}},
    y tick label style = {/pgf/number format/1000 sep={\,}},
    cycle list name    = interpColApproxPlotlist
  ]
  
    \foreach \y in {1, 2, 3, 4}{
      \addplot+ [restrict x to domain=0:20] table[x index = 0, y index = \y] {\tableINPUT};
    }

    \node[anchor=north east, font=\small] at (rel axis cs:1,.97) {Column-based $\BYt \approx \BYangle$ (phasor angle)};
  \end{semilogyaxis}
\end{tikzpicture}%
  \tikzexternaldisable%
\\
    \caption{Relative errors for $K=1,2,\ldots,20$ interpolatory matrix approximations $\BYs$ and $\BYt$ of the phasor angle data $\BYangle.$}
    \label{fig:interpApprox_angle}
\end{subfigure}

\caption{Relative errors for rank $K=1,2\ldots,20$ interpolatory matrix approximations $\BYs$ and $\BYt$ of the matrices $\BYvoltage,\BYangle\in\R^{68\times 6000}$ containing $60$\,s worth of voltage magnitude and phasor angle data generated using the $68$-bus, $16$-machine NETSNYPS test system.
The size of the column-based data prevents the use of MILP in the column-based approximation.}
\label{fig:interpApprox}
\end{figure}
%%%%%%%%%%%%%%%%%%%%%%%%%%%%%%%%%%%%%%%%%%%%%%%%%%%%%%%%%%%%%%%%%

The code and data for reproducing the numerical tests presented in this manuscript are available at~\cite{supRei25}.
All numerical tests were performed on a MacBook Air with 8 gigabytes of RAM and an Apple M2 processor running macOS Sequoia version 15.2 with MATLAB 23.2.0.2515942 (R2023b) Update~7.
We now demonstrate the ability of the row- and column-based IDs~\eqref{eq:interpRowApprox} and~\eqref{eq:interpColApprox} to reduce the dimension of PMU data matrices using different strategies for selecting the pilots $\CS$ and $\CT$.\ \ 
We compare the following selection strategies.
\begin{description}
    \item[DEIM] is Algorithm~\ref{alg:DEIM} from~\cite{ChaS10,Baretal04}.
    \item[QDEIM] is the QDEIM variant of DEIM from~\cite{DrmG16}.
    \item[MILP] selects the indices by solving a mixed-integer linear program (MILP) that minimizes the maximum absolute pairwise cosine similarity among the selected rows/columns, thereby favoring subsets whose vectors are nearly pairwise orthogonal.
    This approach is modeled after the pilot PMU selection strategy proposed in Section~II of~\cite{XieCK14}.\footnote{We note that~\cite{XieCK14} does not explicitly formulate the pilot selection strategy as an MILP; rather, the guiding principle is to choose pilot PMUs such that the cosine similarity among the selected datastreams is close to zero, i.e., the corresponding datastreams are as nearly orthogonal as possible. The specific implementation used to carry out this selection is not detailed therein.} 
    The program is solved using MATLAB's \texttt{intlinprog} command.
    The particular formulation of the MILP and the associated objective function are provided in the Appendix.
    \item[RAND] is a random selection (MATLAB's \texttt{randi} command).
\end{description}
These strategies are applied to matrices $\BY$ of PMU data to identify the sets $\CS$ and $\CT$ with $K$ indices. Then, matrices $\BZs$ and $\BWt$ are computed from~\eqref{eq:lsSol}, and used to form rank-$K$ IDs $\BYs$ and $\BYt$ of the matrix $\BY$ according to~\eqref{eq:interpRowApprox} and~\eqref{eq:interpColApprox}.

To assess the quality of the reduction, we compute the relative errors induced by $\BYs$ and $\BYt$ in the matrix $2$-norm, and compare this against the relative best rank-$k$ approximation error $\sigma_{k+1}/\sigma_1$ from the SVD.\ \ 
We also compute the associated error factors $\etaBs$ and $\etaBt$ for each selection strategy, although we do not report the values of these factors here; they are available in the accompanying code package~\cite{supRei25}.

We test the efficacy of our IDs on synthetic PMU data generated from transient simulations of the NETSNYPS $68$-bus, $16$-machine test system~\cite{PC05}. We perform the simulations with
MATLAB's Power Systems Toolbox (PST)~\cite{chow1992toolbox}.
This setup is similar to that used in~\cite{XieCK14} and~\cite{LiWC18}.
These data correspond to dynamic voltage waveforms and ignore the internal processing mechanism of a PMU, which is manufacturer-dependent.
To mimic realistic operating conditions,
following the setup of~\cite{XieCK14}, zero-mean Gaussian noise is added to all of our synthetically generated PMU data, so that the signal-to-noise ratio (SNR) is 92~dB. 
This noise complies with the accuracy limit of less than 1\% total vector error (TVE) specified by IEEE Standard C37.118.1~\cite{Mar15}.
The components of a realistic PMU measurement chain can exhibit significantly different behavior during a fault, compared to steady-state operating conditions. These uncertainties are not reflected in our synthetic data, which are obtained from simulations.
Real PMU data may also be corrupted by colored noise and suffer from outliers, whereas our methodology assumes white noise and outlier-free data.
Validation of our proposed methodology on data obtained from a PMU emulator or real-world PMU data is a topic that we plan to consider in future work.

We present results on voltage magnitude and phasor angle data at every bus collected at a sampling rate of $100$\,Hz over a $60$\,s window. These data are organized into a pair of $68\times 6000$ dimensional matrices $\BYvoltage$ and $\BYangle$ for voltage magnitudes and phasor angles. \ \ 
After $30$\,s of the simulation, a three-phase line fault is applied between buses $28$ and $29$ and cleared $0.2$\,s later.
For the column-based IDs~\eqref{eq:interpColApprox}, we do not employ the MILP-based selection because the matrices required for solving the program do not fit in RAM.

Figure~\ref{fig:interpApprox} shows
the relative errors in the $2$-norm.
We compute rank $K=1,2,\ldots,20$ row- and column-based IDs for the PMU data matrices $\BYvoltage$ and $\BYangle$ using the selection strategies outlined above and report the relative errors in Figures~\ref{fig:interpApprox_voltage} and~\ref{fig:interpApprox_angle}.
For both types of measurement data, the DEIM- and QDEIM-based IDs give results on par with those of the SVD for each rank $K$.\ \ 
For the voltage magnitude data, the MILP-based IDs perform poorly after an initial reduction for small $K$; the approximation errors oscillate as $K$ increases.
For $K\leq 10$, the MILP-based IDs produce approximations of the phasor angle data that are of similar quality to the DEIM- and QDEIM-based IDs.
For $K > 10$, the row index selection by MILP did not converge as \texttt{intlinprog} terminated early after exploring 1\,000\,000 branch-and-bound nodes. Hence, the approximation error plateaus for these values of $K$.

In all cases, the row-based IDs perform better than the column-based ones, as expected, since the column-based approximations have more indices to choose from ($6000$ vs. $68$).
These results demonstrate that, when combined with a reliable pilot selection strategy, IDs are an effective tool for reducing the dimension of various types of PMU data. 

In the interest of space, we do not report the wall-clock times for the different pilot selection strategies and instead comment on these results for the row-based approximations. The specific timings for each value of $K$ are available in the accompanying code package~\cite{supRei25}.
The DEIM- and QDEIM-based row indices are all computed in less than $1$ second. For the voltage magnitude data, the MILP-based selections are computed in the range of $5$--$15$ seconds for larger values of $K$.  
For the phasor angle data, the MILP-based selection requires significantly more time: for $K=5$, the selection takes approximately $70$ seconds; for $K\geq 8$, this selection takes more than $1\,000$~seconds. 
% SECITON 4
%%%%%%%%%%%%%%%%%%%%%%%%%%%%%%%%%%%%%%%%%%%%%%%%%%%%%%%%%%%%%%%%%%%
\section{Data-driven monitoring with ID-DEIM}
\label{sec:eventMonitoring}
%%%%%%%%%%%%%%%%%%%%%%%%%%%%%%%%%%%%%%%%%%%%%%%%%%%%%%%%%%%%%%%%%%%
Several works have proposed using changes in the low-dimensional subspace spanned by streaming PMU data to detect and localize system events in real time; see, e.g.,~\cite{WanZZ11, Liuetal15, RafLLM16, XieCK14, LiWC18, KonFYY21, WanEtal19}.
Here, we propose an offline-online data-driven framework for real-time monitoring based on the IDs presented in Section~\ref{sec:lowRankPMU} 
and the DEIM index selection algorithm described in Section~\ref{sec:DEIM}.
The proposed framework builds upon the online monitoring algorithm presented in Section~II of~\cite{XieCK14}, insofar as only a reduced number of pilot PMU data streams are used to monitor the network.
Any non-pilot datastreams in the network can be recovered from these pilots, effectively reducing the dimension of the streaming PMU data.

In contrast to the work of~\cite{XieCK14}, our algorithm views this dimension reduction through the lens of IDs~\eqref{eq:interpRowApprox}, enabling the use of the interpolatory error bound $\etaBs \sigma_{K+1}$ in~\eqref{eq:oneSidedInterpErrorBound}.
This perspective yields a few key operational benefits:
\begin{itemize}
    \item Offline, DEIM is applied to select pilot streams $\CS$, increasing the number of pilots $K$ until the bound $\etaBs \sigma_{K+1}$ is less than a user-specified tolerance. 
    \item Online, $\etaBs\sigma_{K+1}$ serves as an \emph{estimator} of the interpolatory reconstruction error; deterioration of this estimate suggests a change in the operating condition of the network, and can be used as a simple ``tripwire'' for detecting disturbances.
    \item Following such a detection, the DEIM algorithm is applied to the transient system response due to the disturbance to localize the source of the event purely from data, with high accuracy.
\end{itemize}
Figure~\ref{fig:flowchart} presents a flowchart depicting the two-stage online-offline workflow of the proposed method.
Numerical tests are interspersed throughout this section to illustrate the proposed framework.

\begin{figure}[t]
\centering
\newcommand{\flowchartfontsize}{\plotfontsize}
\begin{tikzpicture}[
    x=1cm,
    y=1cm,
    font=\flowchartfontsize,
    >=Latex,
    line/.style={-Latex, line width=0.85pt, draw=black!80},
    loopline/.style={line width=0.85pt, draw=black!80}, % CHANGED: no-arrow style for return loops
    hdr/.style={font=\bfseries\flowchartfontsize, text=black},
    badge/.style={circle, minimum size=5mm, inner sep=0pt, font=\bfseries\flowchartfontsize, text=white},
    boxbase/.style={
        draw,
        rounded corners=2pt,
        line width=0.9pt,
        fill=white,
        align=center,
        inner sep=3pt,
        minimum height=7.5mm
    },
    smallboxbase/.style={
        draw,
        rounded corners=2pt,
        line width=0.9pt,
        fill=white,
        align=center,
        inner sep=2.5pt,
        minimum height=7mm
    },
    diamondbase/.style={
        diamond,
        aspect=2.2,
        draw,
        line width=0.9pt,
        align=center,
        inner sep=1pt
    }
]

% -------------------------------------------------
% COLORS
% -------------------------------------------------
\definecolor{myteal}{RGB}{17,125,128}
\definecolor{myblue}{RGB}{33,103,204}
\definecolor{myorange}{RGB}{227,117,22}

% -------------------------------------------------
% WIDTH / LAYOUT KNOBS
% -------------------------------------------------
\def\W{8.5}        % overall panel width  <-- main width knob
\def\HH{0.72}       % header height
\def\CX{4.25}      % x-location of main vertical flow
\def\RX{7.5}        % x-location of right-side loop boxes

\def\MainA{4.7}     % common main box width
\def\MainB{5.0}     % slightly wider main box width
\def\MainC{6.0}     % compact main box width
\def\SmallW{1.55}   % right-side loop box width

% -------------------------------------------------
% PANEL HEIGHT / SPACING KNOBS
% -------------------------------------------------
\def\PanelOneH{7.5}    % CHANGED: height of panel 1
\def\PanelTwoH{6.60}    % CHANGED: taller panel 2
\def\PanelThreeH{3.4}  % CHANGED: taller panel 3

% =================================================
% PANEL 1 : OFFLINE
% =================================================
\begin{scope}[shift={(0,0)}]

    % panel background
    \fill[rounded corners=4pt, fill=myteal!4]
        (0,0) rectangle (\W,-\PanelOneH);
    
    \begin{scope}
        \clip[rounded corners=4pt] (0,0) rectangle (\W,-\PanelOneH);
        \fill[myteal!10] (0,0) rectangle (\W,-\HH);
    \end{scope}
    
    \draw[rounded corners=4pt, draw=myteal, line width=1pt]
        (0,0) rectangle (\W,-\PanelOneH);
    
    \draw[myteal, line width=1pt] (0,-\HH) -- (\W,-\HH);
    
    % header
    \node[badge, fill=myteal] at (0.45,-0.36) {1};
    \node[hdr, anchor=west] at (0.92,-0.36)
        {Offline: Adaptive DEIM-based Training of Pilots};

    % nodes
    \node[boxbase, draw=myteal, text width=\MainA cm] (train) at (\CX,-1.3)
        {Collect ambient training data $\BYtrain$};

    \node[boxbase, draw=myteal, text width=\MainA cm] (svd) at (\CX,-2.5)
        {Compute dominant left singular vectors $\BU_K$ and singular values $\BSigma_{K}$ of $\BYtrain$};

    \node[boxbase, draw=myteal, text width=\MainB cm] (deim) at (\CX,-3.70)
        {Process $\Bu_{k}$ via DEIM to select $k$-th pilot stream
        $\CS_k = \CS_{k-1}\cup\{s_k\}$};

    \node[diamondbase, draw=myteal, fill=myteal!8, text width=2.15cm] (tol) at (\CX,-5.2)
        {$\eta_S\,\sigma_{k+1}\le \tau$?};

    \node[smallboxbase, draw=myteal, text width=\SmallW cm] (add) at (\RX,-5.2)
        {$k\gets k+1$};

    \node[boxbase, draw=myteal, text width=\MainB cm] (prep) at (\CX,-6.8)
        {Set $\CS=\CS_k$ and compute $\BZs$ from\\
        $\displaystyle \min_{\BZ} \|\BYtrain-\BZ\BYtrain(\CS,:)\,\|_{\frob}$};

    % arrows
    \draw[line] (train) -- (svd);
    \draw[line] (svd) -- (deim);
    \draw[line] (deim) -- (tol);

    \draw[line] (tol.east) -- node[above] {No} (add.west);

    % CHANGED: return loop now points back into the DEIM box
    \draw[loopline] (add.north) |- ([xshift=1.5mm]deim.east);
    \draw[line] ([xshift=1.5mm]deim.east) -- (deim.east);

    \draw[line] (tol.south) -- node[right] {Yes} (prep.north);

\end{scope}

% =================================================
% PANEL 2 : ONLINE
% =================================================
% CHANGED: shifted down to avoid overlap with longer Panel 1
\begin{scope}[shift={(0,-7.75)}]

    % panel background
    \fill[rounded corners=4pt, fill=myblue!4]
        (0,0) rectangle (\W,-\PanelTwoH);
    
    \begin{scope}
        \clip[rounded corners=4pt] (0,0) rectangle (\W,-\PanelTwoH);
        \fill[myblue!10] (0,0) rectangle (\W,-\HH);
    \end{scope}
    
    \draw[rounded corners=4pt, draw=myblue, line width=1pt]
        (0,0) rectangle (\W,-\PanelTwoH);
    
    \draw[myblue, line width=1pt] (0,-\HH) -- (\W,-\HH);

    % header
    \node[badge, fill=myblue] at (0.45,-0.36) {2};
    \node[hdr, anchor=west] at (0.92,-0.36)
        {Online Monitoring and Detection};

    % nodes
    % CHANGED: increased vertical spacing between steps
    \node[boxbase, draw=myblue, text width=5.5cm] (obs) at (\CX,-1.40)
        {Collect streaming PMU data \\$\BYo(s_{k},\,j)$ at time $t_{j}$ from pilots $s_{k}\in\CS$};

    \node[boxbase, draw=myblue, text width=\MainC cm] (recon) at (\CX,-2.75)
        {Reconstruct non-pilot streams $i\not \in \CS$\\[.05cm]
        $\BYo(i,\,j)\approx\sum_{k=1}^K \BZs(i, k) \BYo(s_k,\,j)$};

    \node[diamondbase, draw=myblue, fill=myblue!8, text width=2.55cm] (trip) at (\CX,-4.45)
        {$\mathrm{error}>\theta\,\etaBs\,\sigma_{K+1}$?};

    \node[smallboxbase, draw=myblue, text width=1.65cm] (cont) at (\RX,-4.45)
        {Continue monitoring};

    \node[boxbase, draw=myblue, fill=myblue!10, line width=1.1pt, text width=2.55cm] (event) at (\CX,-6.05)
        {\bfseries \emph{Event detected}};

    % \node[smallboxbase, draw=myblue, text width=1.65cm] (retrain) at (\RX,-5.95)
    %     {Proceed to re-trianing};

    % \draw[line] (event) -- (retrain);

    % arrows
    \draw[line] (obs) -- (recon);
    \draw[line] (recon) -- (trip);

    \draw[line] (trip.east) -- node[above] {No} (cont.west);

    % CHANGED: return loop now points back into the streaming-data box
    \draw[loopline] (cont.north) |- ([xshift=1.5mm]obs.east);
    \draw[line] ([xshift=1.5mm]obs.east) -- (obs.east);

    \draw[line] (trip.south) -- node[right] {Yes} (event.north);

\end{scope}

% =================================================
% PANEL 3 : POST-EVENT
% =================================================
% CHANGED: shifted down to account for taller Panel 2
\begin{scope}[shift={(0,-14.6)}]

    % panel background
    \fill[rounded corners=4pt, fill=myorange!4]
        (0,0) rectangle (\W,-\PanelThreeH);
    
    \begin{scope}
        \clip[rounded corners=4pt] (0,0) rectangle (\W,-\PanelThreeH);
        \fill[myorange!10] (0,0) rectangle (\W,-\HH);
    \end{scope}
    
    \draw[rounded corners=4pt, draw=myorange, line width=1pt]
        (0,0) rectangle (\W,-\PanelThreeH);
    
    \draw[myorange, line width=1pt] (0,-\HH) -- (\W,-\HH);

    % header
    \node[badge, fill=myorange] at (0.45,-0.36) {3};
    \node[hdr, anchor=west] at (0.92,-0.36)
        {DEIM-based Post-event Localization};

    % nodes
    % CHANGED: increased vertical spacing between steps
    \node[boxbase, draw=myorange, text width=5.35cm] (window) at (\CX,-1.4)
        {Collect transient window\\ following the disturbance into $\BY$};

    \node[boxbase, draw=myorange, text width=5.55cm] (loc) at (\CX,-2.75)
        {Apply DEIM to $\BU_K$ of pre-processed $\BY$\\
        and order $K$ candidate buses for localization};

    % arrows
    \draw[line] (window) -- (loc);

\end{scope}

\end{tikzpicture}
\caption{Flowchart of the online and offline workflows of the proposed ID-DEIM event monitoring framework for PMU-based event detection and post-event localization.}
\label{fig:flowchart}
\end{figure}

%%%%%%%%%%%%%%%%%%%%%%%%%%%%%%%%%%%%%%%%%%%%%%%%%%%%%%%%%%%%%%%%%%%
\subsection{Adaptive DEIM-based Training of Pilots}
\label{ss:deimTrain}
%%%%%%%%%%%%%%%%%%%%%%%%%%%%%%%%%%%%%%%%%%%%%%%%%%%%%%%%%%%%%%%%%%%
At this point, we differentiate between \emph{offline} training data $\BYtrain\in\R^{N\times \Ttrain}$ and \emph{online} streaming data $\BYo\in\R^{N\times \To}$.
For simplicity, we assume that $\BYtrain$ and $\BYo$ contain data corresponding to a single measured grid quantity. Different quantities are dealt with by placing them into different PMU data matrices and processing them separately.
In the numerical results of this section, we consider voltage magnitude and phasor angle data.
The positive integers $\Ttrain$, $\To$ dictate the size of the training and online monitoring windows. 
The parameter $\Ttrain$ is tunable; in our tests, we choose $\Ttrain$ to correspond to $\sim 120$\,s worth of data collected during normal (ambient) operating conditions. 
On the other hand, we take $\To$ to be fixed, but not necessarily known \emph{a priori}.

For the online monitoring portion of the algorithm, where possibly the full $\BYo$ is approximated from a few pilot datastreams via an ID $\BYs=\BZs\BRs$, two things need to be computed offline from $\BYtrain$ to formulate $\BYs$:
\begin{enumerate}
    \item[{1.}] The indices $\CS$ corresponding to the pilot datastreams that will form the basis of the ID;
    \item[{2.}] The matrix of weights $\BZs\in\R^{N\times K}$ that encodes how the pilot datastreams in $\CS$ should be combined to reproduce measurements at any non-pilot datastreams.
\end{enumerate}
To choose the pilots in $\CS$, DEIM is applied to the leading left singular vectors of $\BYtrain$ until $\etaBs\sigma_{K+1} \leq \tau$, where $\tau > 0$ is a user-specified error tolerance.
In our tests, for modest values of $\tau$, e.g., $\tau=10^{-1}$, the error bound satisfies $\eta_{\CS}\sigma_{K+1} \leq \tau$ for small values of $K$, e.g., $K\leq 5$.
A reliable heuristic for
tuning
$\tau$ is provided by the singular values of $\BYtrain$, since $\sigma_{K+1}$ is the best rank-$K$ approximation error.
A smaller tolerance $\tau$ necessitates more pilots, giving a tradeoff between reconstruction accuracy and the bandwidth required to communicate across the $K$ pilots.
After the pilot streams $\CS$ are identified, the matrix $\BZs$ is computed by solving the least-squares problem~\eqref{eq:lsSol} with $\BY=\BYtrain$ and $\BR_{\CS}=\BYtrain(\CS,\,:\,)$.

This DEIM-based pilot selection method offers key advantages over other selection approaches.
First, by design, the pilots chosen by DEIM yield small values of the error factor $\etaBs$, and thus the bound in~\eqref{eq:oneSidedInterpErrorBound}.
Because of this, and the fact that we expect the data in $\BYo$ to exist near the same low-dimensional subspace spanned by $\BYtrain$, measurements from the non-pilot datastreams can be approximated from the pilots with high fidelity.
Second, spurious retraining can be avoided by periodically recomputing $\eta_{\CS}\sigma_{K+1}$ for the current set of pilots $\CS$ but using a new batch of online data from all datastreams.
So long as the heuristic ``bound'' $\eta_\CS\sigma_{K+1}$ remains below an acceptable threshold, the current configuration of pilots is accepted. Otherwise, retraining is needed: DEIM is applied to $\BYo$ to select a new set of pilots.

%%%%%%%%%%%%%%%%%%%%%%%%%%%%%%%%%%%%%%%%%%%%%%%%%%%%%%%%%%%%%%%%%%%
\emph{Numerical Tests.}
%%%%%%%%%%%%%%%%%%%%%%%%%%%%%%%%%%%%%%%%%%%%%%%%%%%%%%%%%%%%%%%%%%%
%%%%%%%%%%%%%%%%%%%%%%%%%%%%%%%%%%%%%%%%%%%%%%%%%%%%%%%%%%%%%%%%%%%
\begin{figure}[t!]
\centering
  \tikzexternalenable%
  \tikzsetnextfilename{trainingLegend}%
  \begin{tikzpicture}
  \begin{axis}[%
    hide axis,
    width  = 1mm,
    height = 1mm,
    scale only axis,
    xmin = 0,
    xmax = 1,
    ymin = 0,
    ymax = 1,
    legend columns = 5, 
    legend style   = {
      at     = {(0,0)},
      anchor = center,
      /tikz/every even column/.append style = {column sep = 0.1cm}},
    legend cell align  = {left},
    clip mode          = individual,
    cycle list name    = trainingPlotlist]

    \foreach \y in {1, 2, ..., 5}{
      \addplot+ coordinates{ (0, 0) };
    }
    \plotfontsize
    \addlegendentry{Tol. $\tau$}
    \addlegendentry{DEIM}
    \addlegendentry{QDEIM}
    \addlegendentry{MILP}
    \addlegendentry{RAND}
  \end{axis}
\end{tikzpicture}%
  \tikzexternaldisable%

    \vspace{.5\baselineskip}

\begin{subfigure}[b!]{\linewidth}
    \raggedleft
  \tikzexternalenable%
  \tikzsetnextfilename{trainingRowLebesgueConstants_busV}%
  \begin{tikzpicture}[font = \plotfontsize]
  \pgfplotstableread{graphics/data/training_error_factors_busV.dat}\tableINPUT
  
  \begin{semilogyaxis}[%
    width  = .825\linewidth,
    height = .1\textheight,
    scale only axis,
    grid=both,
    grid style={line width=.1pt, draw=gray!10},
    major grid style={line width=.2pt,draw=gray!25},
    minor tick num=1,
    xtick = {1, 2, 4, 6, 8, 10},
    extra x ticks = {3, 5, 7, 9},
    extra x tick labels={},
    xmin = 1,
    xmax = 10,
    ymin = 0,
    ymax = 1e5,
    xminorticks = true,
    yminorticks = true,
    xlabel = {$K$, number of pilots},
    ylabel = {$\etaBs$},
    ylabel style   = {yshift = -.3em},
    x tick label style = {/pgf/number format/1000 sep={\,}},
    y tick label style = {/pgf/number format/1000 sep={\,}},
    cycle list name    = interpApproxPlotlist
  ]

  \pgfplotsset{cycle list shift = 1}
  
    \foreach \y in {1,2,3,4}{
      \addplot+ [restrict x to domain=0:10] table[x index = 0, y index = \y] {\tableINPUT};
    }

    % \node[anchor=north, font=\small] at (rel axis cs:1,1) {Voltage mag.};
  \end{semilogyaxis}
\end{tikzpicture}%
  \tikzexternaldisable%
\\
    \vspace{.5\baselineskip}
  \tikzexternalenable%
  \tikzsetnextfilename{trainingRowErrorBound_busV}%
  \begin{tikzpicture}[font = \plotfontsize]
  \pgfplotstableread{graphics/data/training_error_bounds_busV.dat}\tableINPUT
  
  \begin{semilogyaxis}[%
    width  = .825\linewidth,
    height = .1\textheight,
    scale only axis,
    grid=both,
    grid style={line width=.1pt, draw=gray!10},
    major grid style={line width=.2pt,draw=gray!25},
    minor tick num=1,
    xtick = {1, 2, 4, 6, 8, 10},
    extra x ticks = {3, 5, 7, 9},
    extra x tick labels={},
    xmin = 1,
    xmax = 10,
    ymin = 0,
    ymax = 1e3,
    xminorticks = true,
    yminorticks = true,
    xlabel = {$K$, number of pilots},
    ylabel = {$\etaBs\,\sigma_{K+1}$},
    ylabel style   = {yshift = -.3em},
    x tick label style = {/pgf/number format/1000 sep={\,}},
    y tick label style = {/pgf/number format/1000 sep={\,}},
    cycle list name    = interpApproxPlotlist
  ]

  \pgfplotsset{cycle list shift = 1}
  
    \foreach \y in {1,2,3,4}{
      \addplot+ [restrict x to domain=0:10]  table[x index = 0, y index = \y] {\tableINPUT};
    }

    \addplot [domain=0:10, samples=2, dashed, black] {5*1e-2};
    \node[anchor=south] at (axis cs:8, 4*1e-2) {$\tau = 5 \times 10^{-2}$};
    
  \end{semilogyaxis}
\end{tikzpicture}%
  \tikzexternaldisable%
\\
    \caption{Evolution of the error factor $\etaBs$ and the upper bound $\etaBs\sigma_{K+1}$ throughout the adaptive training applied to the \emph{voltage magnitude} data.}
    \label{fig:training_voltage}
    \vspace{.5\baselineskip}
\end{subfigure}
\begin{subfigure}[b!]{\linewidth}
    \raggedleft
  \tikzexternalenable%
  \tikzsetnextfilename{trainingRowLebesgueConstants_thetaV}%
  \begin{tikzpicture}[font = \plotfontsize]
  \pgfplotstableread{graphics/data/training_error_factors_thetaV.dat}\tableINPUT
  
  \begin{semilogyaxis}[%
    width  = .825\linewidth,
    height = .1\textheight,
    scale only axis,
    grid=both,
    grid style={line width=.1pt, draw=gray!10},
    major grid style={line width=.2pt,draw=gray!25},
    minor tick num=1,
    xtick = {1, 2, 4, 6, 8, 10},
    extra x ticks = {3, 5, 7, 9},
    extra x tick labels={},
    xmin = 1,
    xmax = 10,
    ymin = 0,
    ymax = 1e6,
    xminorticks = true,
    yminorticks = true,
    xlabel = {$K$, number of pilots},
    ylabel = {$\etaBs$},
    ylabel style   = {yshift = -.3em},
    x tick label style = {/pgf/number format/1000 sep={\,}},
    y tick label style = {/pgf/number format/1000 sep={\,}},
    cycle list name    = interpApproxPlotlist
  ]

  \pgfplotsset{cycle list shift = 1}
  
    \foreach \y in {1,2,3,4}{
      \addplot+ [restrict x to domain=0:10] table[x index = 0, y index = \y] {\tableINPUT};
    }

  \end{semilogyaxis}
\end{tikzpicture}%
  \tikzexternaldisable%
\\
    \vspace{.5\baselineskip}
  \tikzexternalenable%
  \tikzsetnextfilename{trainingRowErrorBound_thetaV}%
  \begin{tikzpicture}[font = \plotfontsize]
  \pgfplotstableread{graphics/data/training_error_bounds_thetaV.dat}\tableINPUT
  
  \begin{semilogyaxis}[%
    width  = .825\linewidth,
    height = .1\textheight,
    scale only axis,
    grid=both,
    grid style={line width=.1pt, draw=gray!10},
    major grid style={line width=.2pt,draw=gray!25},
    minor tick num=1,
    xtick = {1, 2, 4, 6, 8, 10},
    extra x ticks = {3, 5, 7, 9},
    extra x tick labels={},
    xmin = 1,
    xmax = 10,
    ymin = 0,
    ymax = 1e4,
    xminorticks = true,
    yminorticks = true,
    xlabel = {$K$, number of pilots},
    ylabel = {$\etaBs\,\sigma_{K+1}$},
    ylabel style   = {yshift = -.3em},
    x tick label style = {/pgf/number format/1000 sep={\,}},
    y tick label style = {/pgf/number format/1000 sep={\,}},
    cycle list name    = interpApproxPlotlist
  ]

  \pgfplotsset{cycle list shift = 1}
  
    \foreach \y in {1,2,3,4}{
      \addplot+ [restrict x to domain=0:10]  table[x index = 0, y index = \y] {\tableINPUT};
    }

    % \addplot [domain=0:10, samples=2, dashed, black] {5*1e-2};
    % \node[anchor=south] at (axis cs:8.5, 1.5*1e-1) {$\tau = 5 \times 10^{-2}$};

    \addplot [domain=0:10, samples=2, dashed, black] {5*1e-1};
    \node[anchor=south] at (axis cs:8.5, 4*1e-1) {$\tau = 5\times10^{-1}$};
    
  \end{semilogyaxis}
\end{tikzpicture}%
  \tikzexternaldisable%
\\
    \caption{Evolution of the error factor $\etaBs$ and the upper bound $\etaBs\sigma_{K+1}$ throughout the adaptive training applied the \emph{phasor angle} data.}
    \label{fig:training_angle}
    \vspace{\baselineskip}
\end{subfigure}
\vspace{-.5\baselineskip}
\caption{Evolution of the error factors $\etaBs$ and the upper bounds $\etaBs\sigma_{K+1}$ throughout the adaptive training as $K$ pilots are chosen.}
\label{fig:training}
\end{figure}
%%%%%%%%%%%%%%%%%%%%%%%%%%%%%%%%%%%%%%%%%%%%%%%%%%%%%%%%%%%%%%%%%%%
We test the adaptive DEIM-based training of pilot-stream configurations on synthetic PMU data from the NETSNYPS $68$-bus, $16$-machine test system.
Two sets of training data containing ambient voltage magnitudes and phasor angles are collected at a rate of $100$\,Hz over $120\,$s.
To mimic ambient operating conditions, the mechanical power and exciter references of the generators are perturbed by zero-mean Gaussian white noise during the simulation.

The adaptive DEIM-based training algorithm is applied to the voltage magnitude and angle training data, separately, to compute pilot configurations for each measurement type.
To observe the evolution of the error factor $\eta_{\CS}$ and the bound $\eta_{\CS}\sigma_{K+1}$ throughout the training procedure, we run DEIM for $K = 10$ iterations.
We compare the DEIM-based training with pilot configurations computed from $\BYtrain$ using the QDEIM, MILP, and RAND selection schemes described in Section~\ref{ss:interpNumerics}.
Because these other schemes are not inherently iterative like DEIM, we run them repeatedly for each fixed size ${K = 1, 2, \ldots, 10}$ of $\CS$ 
(and thus, unlike DEIM, the resulting index sets need not be nested).

We report our results in Figure~\ref{fig:training}. The evolution of $\eta_{\CS}$ and $\eta_{\CS}\sigma_{K+1}$ as $K$, the number of pilots, grows for the voltage magnitude and phasor angle training data are shown in Figures~\ref{fig:training_voltage} and~\ref{fig:training_angle}.
We observe in both instances that these quantities steadily decrease with $K$ as each new pilot is selected by DEIM and QDEIM.\ \ 
For the MILP- and RAND-based selections, both the error factor and thus the error bound tend to oscillate and generally \emph{increase} as more pilots are added.

%%%%%%%%%%%%%%%%%%%%%%%%%%%%%%%%%%%%%%%%%%%%%%%%%%%%%%%%%%%%%%%%%%%
\subsection{Online Monitoring and Detection Using the Bound~\eqref{eq:oneSidedInterpErrorBound}}
\label{ss:eventDet}
%%%%%%%%%%%%%%%%%%%%%%%%%%%%%%%%%%%%%%%%%%%%%%%%%%%%%%%%%%%%%%%%%%%
After using $\BYtrain$ to determine the $K$ pilot streams in $\CS$ and the matrix $\BZs$ offline, dimension reduction of the online data $\BYo$ is achieved in real time via a rank-$K$ ID $\BYs$.\ \ 
Suppose we want to recover data from one of the non-pilot streams $i\not\in\CS$ at the $j$-th time snapshot, i.e., $\BYo(i,\,j)$.
To recover $\BYo(i,\,j)$, data from each of the $K$ pilot streams $\BYo(s_k,\,j)$, $k = 1, \ldots, K$, are combined according to
\begin{equation}
\label{eq:interpRowApprox_stream}
    \BYo(i,\,j)\approx\BYs(i,\, j) :=\sum_{k=1}^K \BZs(i, k) \BYo(s_k,\,j).
\end{equation}
We emphasize that the datastreams indicated by $\CS$ are chosen based on the offline data $\BYtrain$, but the actual \emph{online} datastreams in $\BYo$ are used to form the reconstruction~\eqref{eq:interpRowApprox_stream}.
Because the training and online data are not the same, $\etaBs\sigma_{K+1}$ (from $\BYtrain$) does not provide a theoretically rigorous upper bound for $|\BYo(i,j)-\BYs(i,j)|$, the online reconstruction error for the $j$-th sample of the non-pilot stream $i\not\in\CS$.
However, the factor $\etaBs\sigma_{K+1}$ can serve as an \emph{estimator} of the online reconstruction error:
because the training data $\BYtrain$ reflect normal operating conditions, we expect the low-dimensional subspaces spanned by $\BYtrain$ and $\BYo$ to be similar.

Suppose now that the actual reconstruction error of the $j$-th sample collected from a non-pilot stream $i\not\in\CS$ satisfies
\begin{equation}
\label{eq:boundViolated}
  \big|\BYo(i,j)-\BYs(i,j)\big|>\theta\, \etaBs\,\sigma_{K+1}
\end{equation}
for a calibration parameter $\theta > 0$. The condition~\eqref{eq:boundViolated} supposes that the \emph{actual} reconstruction error exceeds the \emph{estimated} error provided by the scaled error factor $\theta\,\etaBs\,\sigma_{K+1}$.
This violation suggests a flaw in the assumption that the low-dimensional column space of $\BYo$ is close to that of $\BYtrain$, and thus signals that there has been a fundamental change in the network's operating condition, e.g., due to a disturbance. 

We propose that the deterioration of the error estimate as
in~\eqref{eq:boundViolated} be used as a simple mechanism for detecting changes
to the network’s operating conditions.
The parameter $\theta > 0$ is a calibration multiplier that rescales the error estimator $\etaBs\,\sigma_{K+1}$ and is used to balance the possibility of false alarms against missed disturbances.
Taking $\theta > 1$ increases the chance of false negatives (FN: a change in the network occurs but goes unnoticed), whereas taking $\theta < 1$ will increase the chance of false positives (FP: a change in the network is detected where none has occurred).
A practical way to choose $\theta$ is to compute the time-averaged value of the ratio $|\BYo(i,j)-\BYs(i,j)|/(\etaBs\,\sigma_{K+1})$ for a non-pilot stream $i\not\in\CS$ during nominal operating conditions.
Setting $\theta$ to this ratio would likely trigger an FP.\ \ Therefore, depending on the user's tolerance for FPs, $\theta$ can be chosen to be a value less than the computed ratio.

Necessarily, checking~\eqref{eq:boundViolated} requires continuously monitoring the interpolatory reconstruction error at some non-pilot data\-streams $i\not\in\CS$.\ \ 
These non-pilot data\-streams can be chosen based on the geography of the network, e.g., they may be collected from PMUs that are geographically distributed, to monitor for disturbances that are localized to a subnetwork.
For our tests, we let the DEIM-based training procedure continue to run for $M$ more iterations after $\etaBs\sigma_{K+1} \leq \tau$ is satisfied, for some positive integer $M$. We use these $M$ additional indices as the monitored non-pilot datastreams.
Lastly, we mention that a similar detection mechanism based on the deterioration of the pilot-based reconstruction error is proposed in Section~III.B of~\cite{XieCK14}. By comparison, our detector~\eqref{eq:boundViolated} is grounded in the theoretical upper bound on the interpolatory approximation error~\eqref{eq:oneSidedInterpErrorBound}.

%%%%%%%%%%%%%%%%%%%%%%%%%%%%%%%%%%%%%%%%%%%%%%%%%%%%%%%%%%%%%%%%%%
\emph{Numerical Tests.}
%%%%%%%%%%%%%%%%%%%%%%%%%%%%%%%%%%%%%%%%%%%%%%%%%%%%%%%%%%%%%%%%%%
We investigate the proposed online monitoring algorithm with two numerical tests.
First, we apply the algorithm to the voltage magnitude and phasor angle measurement data collected from the $60\,$s online simulation scenario of the NETSNYPS 68-bus, 16-machine test system from Section~\ref{ss:interpNumerics} during which a three-phase fault is applied to the line connecting buses $28$ and $29$ after $30\,$s.
To select the pilot streams for each measurement type, we re-run the adaptive DEIM-based training algorithm on the $120\,$s of voltage magnitude and phasor angle training data from Section~\ref{ss:deimTrain} using $\tau = 5\times 10^{-2}$ for the voltage data and $\tau = 5\times 10^{-1}$ for the phasor angle data.
(We use different tolerances for the two measurement types because the singular values of the voltage magnitude data decay more rapidly than those of the phasor angle data.)

The adaptive DEIM-based training algorithm selects $K = 5$ pilot streams for the voltage magnitude data and $K = 3$ pilot streams for the phasor angle data, both satisfying the prescribed error tolerances.
For comparison, we run MILP to select $K = 5$ voltage magnitude pilot streams and $K = 3$ phasor angle pilot streams.  Tables~\ref{tab:pilots_voltage} and~\ref{tab:pilots_angle} report the pilot configurations for the voltage magnitudes and phasor angle data, along with the associated error bounds $\etaBs\,\sigma_{K+1}$. 
For the same fixed $K$, the DEIM-based pilots produce an error bound that is roughly an order of magnitude less than the bound for the MILP-based pilots. 
For this experiment, we use $M = 2$ non-pilot streams to monitor for each measurement type: these are the next two pilots identified by DEIM after $K = 5$ and $K=3$. The non-pilots correspond to the voltage magnitudes at buses $63$ and $67$, and the phasor angles at buses $61$ and $68$.

%%%%%%%%%%%%%%%%%%%%%%%%%%%%%%%%%%%%%%%%%%%%%%%%%%%%%%%%%%%%%%%%%%%
\begin{table}[t!]
  \caption{$K=5$ voltage magnitude pilot datastreams as chosen by DEIM and MILP with a tolerance of $\tau=5\times 10^{-2}$, and values of the corresponding error estimator in~\eqref{eq:oneSidedInterpErrorBound}.}
  \label{tab:pilots_voltage}
  
  \centering
  \vspace{.5\baselineskip}

  \begin{tabular}{lrrrrrr}
    \hline\noalign{\smallskip}
      & \multicolumn{1}{c}{$s_1$}
      & \multicolumn{1}{c}{$s_2$}
      & \multicolumn{1}{c}{$s_3$}
      & \multicolumn{1}{c}{$s_4$}
      & \multicolumn{1}{c}{$s_5$}
      & \multicolumn{1}{c}{$\etaBs\sigma_{6}$}\\
    \noalign{\smallskip}\hline\noalign{\smallskip}
    DEIM
      & $48$
      & $61$
      & $59$
      & $55$
      & $50$
      & $4.6488\texttt{e-}2$\\
    MILP 
      & $4$
      & $10$
      & $16$
      & $50$
      & $61$
      & $4.8467\texttt{e-}1$\\

\noalign{\smallskip}\hline\noalign{\smallskip}
  \end{tabular}
\end{table}
%%%%%%%%%%%%%%%%%%%%%%%%%%%%%%%%%%%%%%%%%%%%%%%%%%%%%%%%%%%%%%%%%%%

%%%%%%%%%%%%%%%%%%%%%%%%%%%%%%%%%%%%%%%%%%%%%%%%%%%%%%%%%%%%%%%%%%%
\begin{table}[t!]
  \caption{$K=3$ phasor angle pilot datastreams as chosen by DEIM and MILP with a tolerance of $\tau=5\times 10^{-1}$, and values of the corresponding error estimator in~\eqref{eq:oneSidedInterpErrorBound}.}
  \label{tab:pilots_angle}
  \centering
  \vspace{.5\baselineskip}
  \begin{tabular}{lrrrl}
    \hline\noalign{\smallskip}
      & \multicolumn{1}{c}{$s_1$}
      & \multicolumn{1}{c}{$s_2$}
      & \multicolumn{1}{c}{$s_3$}
      & \multicolumn{1}{c}{$\etaBs\sigma_{4}$}\\
    \noalign{\smallskip}\hline\noalign{\smallskip}
    DEIM
      & $66$
      & $39$
      & $57$
      & $3.3689\texttt{e-}1$\\
    MILP 
      & $36$
      & $45$
      & $65$
      & $4.1213\texttt{e}0$\\
\noalign{\smallskip}\hline\noalign{\smallskip}
  \end{tabular}
\end{table}
%%%%%%%%%%%%%%%%%%%%%%%%%%%%%%%%%%%%%%%%%%%%%%%%%%%%%%%%%%%%%%%%%%%

%%%%%%%%%%%%%%%%%%%%%%%%%%%%%%%%%%%%%%%%%%%%%%%%%%%%%%%%%%%%%%%%%%%
\begin{figure}[t!]
    \centering
  \tikzexternalenable%
  \tikzsetnextfilename{busLegend}%
  \begin{tikzpicture}
  \begin{axis}[%
    hide axis,
    width  = 1mm,
    height = 1mm,
    scale only axis,
    xmin = 0,
    xmax = 1,
    ymin = 0,
    ymax = 1,
    legend columns = 3, 
    legend style   = {
      at     = {(0,0)},
      anchor = center,
      /tikz/every even column/.append style = {column sep = 0.2cm}},
    legend cell align  = {left},
    clip mode          = individual,
    cycle list name    = busPlotlist]

    \foreach \y in {1, 2, 3, 4, 5}{
      \addplot+ coordinates{ (0, 0) };
    }
    \plotfontsize
    \addlegendentry{Original}
    \addlegendentry{DEIM}
    \addlegendentry{MILP}
    \addlegendentry{$\etaBs\sigma_{K+1}$ (DEIM)}
    \addlegendentry{$\etaBs\sigma_{K+1}$ (MILP)}
  \end{axis}
\end{tikzpicture}%
  \tikzexternaldisable%

    \vspace{\baselineskip}
    \begin{subfigure}[b]{.48\linewidth}
    \raggedleft
  \tikzexternalenable%
  \tikzsetnextfilename{singlesim_bus63reconstruction_preEvent_busV}%
  \begin{tikzpicture}[font = \plotfontsize]
  \pgfplotstableread{graphics/data/single_sim_bus63reconstruction_pre_busV.dat}\tableINPUT
  
  \begin{axis}[%
    width  = .6\linewidth,
    height = .10\textheight,
    scale only axis,
    xmin = 5,
    xmax = 10,
    ymin = 0.99985,
    ymax = 1.00015,
    yticklabel style={/pgf/number format/.cd,fixed zerofill,precision=4},
    ytick = {0.9999,1.0000,1.0001},
    ytick distance = .0001,
    xminorticks = true,
    xlabel = {$t_i$, seconds},
    ylabel = {voltage mag. (p.u.)},
    ylabel style   = {yshift = -.3em},
    x tick label style = {/pgf/number format/1000 sep={\,}},
    y tick label style = {/pgf/number format/1000 sep={\,}},
    cycle list name    = busPlotlist
  ]

    \foreach \y in {1,2,3}{
      \addplot+ table[x index = 0, y index = \y] {\tableINPUT};
    }    

    \node[anchor=north east, font=\small] at (rel axis cs:1,1.02) {Bus 63};
    
  \end{axis}
\end{tikzpicture}%
  \tikzexternaldisable%
\\
        \vspace{.5\baselineskip}
  \tikzexternalenable%
  \tikzsetnextfilename{singlesim_bus63errors_preEvent_busV}%
  \begin{tikzpicture}[font = \plotfontsize]
  \pgfplotstableread{graphics/data/single_sim_bus63errors_pre_busV.dat}\tableINPUT
  
  \begin{semilogyaxis}[%
    width  = .6\linewidth,
    height = .10\textheight,
    scale only axis,
    xmin = 5,
    xmax = 10,
    ymin = 1e-9,
    ymax = 1e2,
    /pgf/number format/fixed,
    /pgf/number format/precision=6,
    xminorticks = true,
    % yminorticks = true,
    xlabel = {$t_i$, seconds},
    ylabel = {relative error},
    ylabel style   = {yshift = -.3em},
    scaled x ticks = false,
    x tick label style = {/pgf/number format/1000 sep={\,}},
    y tick label style = {/pgf/number format/1000 sep={\,}},
    cycle list name    = busPlotlist
  ]

  \pgfplotsset{cycle list shift = 1}
    \foreach \y in {1,2}{
      \addplot+ table[x index = 0, y index = \y] {\tableINPUT};
    }    

    \addplot [domain=5:10, samples=2, solid, matlaborange, line width = 1.5pt] {0.0424};

    \addplot [domain=5:10, samples=2, solid, matlabgreen, line width = 1.5pt] {31.12};
  \end{semilogyaxis}
\end{tikzpicture}%
  \tikzexternaldisable%
\\
        \vspace{\baselineskip}
    \end{subfigure}
    \begin{subfigure}[b]{.48\linewidth}
    \raggedleft
  \tikzexternalenable%
  \tikzsetnextfilename{singlesim_bus63reconstruction_during_busV}%
  \begin{tikzpicture}[font = \plotfontsize]
  \pgfplotstableread{graphics/data/single_sim_bus63reconstruction_during_busV.dat}\tableINPUT
  
  \begin{axis}[%
    width  = .6\linewidth,
    height = .10\textheight,
    scale only axis,
    xmin = 29,
    xmax = 31,
    ymin = 0.8400,
    ymax = 1.0600,
    yticklabel style={/pgf/number format/.cd,fixed zerofill,precision=4},
    ytick = {0.8600,0.9500,1.0400},
    xminorticks = true,
    xlabel = {$t_i$, seconds},
    x tick label style = {/pgf/number format/1000 sep={\,}},
    y tick label style = {/pgf/number format/1000 sep={\,}},
    cycle list name    = busPlotlist,
  ]

    \foreach \y in {1,2,3}{
      \addplot+ table[x index = 0, y index = \y] {\tableINPUT};
    }    

    \node[anchor=north east, font=\small] at (rel axis cs:.95,1.02) {Bus 63};
    
  \end{axis}
\end{tikzpicture}%
  \tikzexternaldisable%
\\
        \vspace{.5\baselineskip}
  \tikzexternalenable%
  \tikzsetnextfilename{singlesim_bus63errors_during_busV}%
  \begin{tikzpicture}[font = \plotfontsize]
  \pgfplotstableread{graphics/data/single_sim_bus63errors_during_busV.dat}\tableINPUT
  
  \begin{semilogyaxis}[%
    width  = .6\linewidth,
    height = .10\textheight,
    scale only axis,
    xmin = 29,
    xmax = 31,
    ymin = 1e-9,
    ymax = 1e2,,
    /pgf/number format/fixed,
    /pgf/number format/precision=6,
    xminorticks = true,
    % yminorticks = true,
    xlabel = {$t_i$, seconds},
    % clip marker paths=true,
    % ylabel = {$\relerr(t_i)$},
    % ylabel style   = {yshift = -.3em},
    scaled x ticks = false,
    x tick label style = {/pgf/number format/1000 sep={\,}},
    y tick label style = {/pgf/number format/1000 sep={\,}},
    cycle list name    = busPlotlist
  ]

  \pgfplotsset{cycle list shift = 1}
    \foreach \y in {1,2}{
      \addplot+ table[x index = 0, y index = \y] {\tableINPUT};
    }    

    \addplot [domain=29:31, samples=2, solid, matlaborange, line width = 1.5pt] {0.0424};

    \addplot [domain=29:31, samples=2, solid, matlabgreen, line width = 1.5pt] {31.12};
    
    % \node[overlay, matlabred] at (axis cs:30.2,0.0424) {\Large$\times$};
    % \node[overlay, circle, fill=matlaborange, inner sep=2.5pt] at (axis cs:30,0.0424) {};
    \node[overlay, text=black, font=\Large\bfseries] at (axis cs:30.2,0.0424) {$\times$};
  \end{semilogyaxis}
\end{tikzpicture}%
  \tikzexternaldisable%
\\
        \vspace{\baselineskip}
    \end{subfigure}\\
    \begin{subfigure}[b]{.48\linewidth}
    \raggedleft
  \tikzexternalenable%
  \tikzsetnextfilename{singlesim_bus67reconstruction_preEvent_busV}%
  \usetikzlibrary{spy}
\begin{tikzpicture}[font = \plotfontsize, spy using outlines={rectangle, magnification=4, size=1.5cm, connect spies}]
\pgfplotstableread{graphics/data/single_sim_bus67reconstruction_pre_busV.dat}\tableINPUT
  
  \begin{axis}[%
    width  = .6\linewidth,
    height = .10\textheight,
    scale only axis,
    xmin = 5,
    xmax = 10,
    ymin = 0.99985,
    ymax = 1.00015,
    yticklabel style={/pgf/number format/.cd,fixed zerofill,precision=4},
    ytick = {0.9999,1.0000,1.0001},
    ytick distance = .0001,
    xminorticks = true,
    xlabel = {$t_i$, seconds},
    ylabel = {voltage mag. (p.u.)},
    ylabel style   = {yshift = -.3em},
    x tick label style = {/pgf/number format/1000 sep={\,}},
    y tick label style = {/pgf/number format/1000 sep={\,}},
    cycle list name    = busPlotlist
  ]
  
    \foreach \y in {1,2,3}{
      \addplot+ table[x index = 0, y index = \y] {\tableINPUT};
    }    

        \node[anchor=north east, font=\small] at (rel axis cs:1,1.02) {Bus 67};

  \end{axis}
\end{tikzpicture}%
  \tikzexternaldisable%
\\
        \vspace{.5\baselineskip}
  \tikzexternalenable%
  \tikzsetnextfilename{singlesim_bus67errors_preEvent_busV}%
  \begin{tikzpicture}[font = \plotfontsize]
  \pgfplotstableread{graphics/data/single_sim_bus67errors_pre_busV.dat}\tableINPUT
  
  \begin{semilogyaxis}[%
    width  = .6\linewidth,
    height = .10\textheight,
    scale only axis,
    xmin = 5,
    xmax = 10,
    ymin = 1e-9,
    ymax = 1e2,
    /pgf/number format/fixed,
    /pgf/number format/precision=6,
    xminorticks = true,
    % yminorticks = true,
    xlabel = {$t_i$, seconds},
    ylabel = {relative error},
    ylabel style   = {yshift = -.3em},
    scaled x ticks = false,
    x tick label style = {/pgf/number format/1000 sep={\,}},
    y tick label style = {/pgf/number format/1000 sep={\,}},
    cycle list name    = busPlotlist
  ]

  \pgfplotsset{cycle list shift = 1}
    \foreach \y in {1,2}{
      \addplot+ table[x index = 0, y index = \y] {\tableINPUT};
    }    

    \addplot [domain=5:10, samples=2, solid, matlaborange, line width = 1.5pt] {0.0424};
    % \node[anchor=north] at (axis cs:8, 0.0424) {{$\eta_{\CS_5}\,\sigma_{6}= 4.239\times 10^{-2}$ (DEIM)}};

    \addplot [domain=5:10, samples=2, solid, matlabgreen, line width = 1.5pt] {31.12};
    % \node[anchor=north] at (axis cs:8, 1.18) {{$\eta_{\CS_5}\,\sigma_{6}= 1.188\times 10^{0}$ (MILP)}};
  \end{semilogyaxis}
\end{tikzpicture}%
  \tikzexternaldisable%
\\
        \vspace{\baselineskip}
    \end{subfigure}
    \begin{subfigure}[b]{.48\linewidth}
    \raggedleft
  \tikzexternalenable%
  \tikzsetnextfilename{singlesim_bus67reconstruction_during_busV}%
  \usetikzlibrary{spy}
\begin{tikzpicture}[font = \plotfontsize, spy using outlines={rectangle, magnification=4, size=1.5cm, connect spies}]
\pgfplotstableread{graphics/data/single_sim_bus67reconstruction_during_busV.dat}\tableINPUT
  
  \begin{axis}[%
    width  = .6\linewidth,
    height = .10\textheight,
    scale only axis,
    xmin = 29,
    xmax = 31,
    ymin = 0.7900,
    ymax = 1.080,
    yticklabel style={/pgf/number format/.cd,fixed zerofill,precision=4},
    ytick = {0.8400,0.9500,1.0600},
    xminorticks = true,
    xlabel = {$t_i$, seconds},
    x tick label style = {/pgf/number format/1000 sep={\,}},
    y tick label style = {/pgf/number format/1000 sep={\,}},
    cycle list name    = busPlotlist
  ]

    \foreach \y in {1,2,3}{
      \addplot+ table[x index = 0, y index = \y] {\tableINPUT};
    }    

    \node[anchor=north east, font=\small] at (rel axis cs:1,1.02) {Bus 67};

  \end{axis}
\end{tikzpicture}%
  \tikzexternaldisable%
\\
        \vspace{.5\baselineskip}
  \tikzexternalenable%
  \tikzsetnextfilename{singlesim_bus67errors_during_busV}%
  \begin{tikzpicture}[font = \plotfontsize]
  \pgfplotstableread{graphics/data/single_sim_bus67errors_during_busV.dat}\tableINPUT
  
  \begin{semilogyaxis}[%
    width  = .6\linewidth,
    height = .10\textheight,
    scale only axis,
    xmin = 29,
    xmax = 31,
    ymin = 1e-9,
    ymax = 1e2,
    /pgf/number format/fixed,
    /pgf/number format/precision=6,
    xminorticks = true,
    % yminorticks = true,
    xlabel = {$t_i$, seconds},
    % ylabel = {$\relerr(t_i)$},
    % ylabel style   = {yshift = -.3em},
    scaled x ticks = false,
    x tick label style = {/pgf/number format/1000 sep={\,}},
    y tick label style = {/pgf/number format/1000 sep={\,}},
    cycle list name    = busPlotlist
  ]

  \pgfplotsset{cycle list shift = 1}
    \foreach \y in {1,2}{
      \addplot+ table[x index = 0, y index = \y] {\tableINPUT};
    }    

    \addplot [domain=29:31, samples=2, solid, matlaborange, line width = 1.5pt] {0.0424};

    \addplot [domain=29:31, samples=2, solid, matlabgreen, line width = 1.5pt] {31.12};
    \node[overlay, text=black, font=\Large\bfseries] at (axis cs:30.02,0.0424) {$\times$};
  \end{semilogyaxis}
\end{tikzpicture}%
  \tikzexternaldisable%
\\
        \vspace{\baselineskip}
    \end{subfigure}

\vspace*{-\baselineskip}
    \caption{Interpolatory reconstruction and true data for the non-pilot voltage magnitude datastreams at bus $63$ (top) and bus $67$ (bottom) before and during a three-phase fault of the line between buses $28$ and $29$. The black marker $\times$ indicates the precise instance at which the error estimator is violated according to~\eqref{eq:boundViolated} with $\theta = 1$.}
    \label{fig:singleSim_voltage}
\end{figure}
%%%%%%%%%%%%%%%%%%%%%%%%%%%%%%%%%%%%%%%%%%%%%%%%%%%%%%%%%%%%%%%%%%%

%%%%%%%%%%%%%%%%%%%%%%%%%%%%%%%%%%%%%%%%%%%%%%%%%%%%%%%%%%%%%%%%%%%
\begin{figure}[t!]
    \centering
  \tikzexternalenable%
  \tikzsetnextfilename{busLegend}%
  \begin{tikzpicture}
  \begin{axis}[%
    hide axis,
    width  = 1mm,
    height = 1mm,
    scale only axis,
    xmin = 0,
    xmax = 1,
    ymin = 0,
    ymax = 1,
    legend columns = 3, 
    legend style   = {
      at     = {(0,0)},
      anchor = center,
      /tikz/every even column/.append style = {column sep = 0.2cm}},
    legend cell align  = {left},
    clip mode          = individual,
    cycle list name    = busPlotlist]

    \foreach \y in {1, 2, 3, 4, 5}{
      \addplot+ coordinates{ (0, 0) };
    }
    \plotfontsize
    \addlegendentry{Original}
    \addlegendentry{DEIM}
    \addlegendentry{MILP}
    \addlegendentry{$\etaBs\sigma_{K+1}$ (DEIM)}
    \addlegendentry{$\etaBs\sigma_{K+1}$ (MILP)}
  \end{axis}
\end{tikzpicture}%
  \tikzexternaldisable%

    \vspace{\baselineskip}
    \begin{subfigure}[b]{.48\linewidth}
    \raggedleft
  \tikzexternalenable%
  \tikzsetnextfilename{singlesim_bus61reconstruction_preEvent_thetaV}%
  \usetikzlibrary{spy}
\begin{tikzpicture}[font = \plotfontsize, spy using outlines={rectangle, magnification=4, size=1.5cm, connect spies}]
\pgfplotstableread{graphics/data/single_sim_bus61reconstruction_pre_thetaV.dat}\tableINPUT
  
  \begin{axis}[%
    width  = .6\linewidth,
    height = .10\textheight,
    scale only axis,
    xmin = 5,
    xmax = 10,
    ymin = 0.3620,
    ymax = 0.3680,
    yticklabel style={/pgf/number format/.cd,fixed zerofill,precision=4},
    ytick = {0.3670,0.3650,0.3630},
    ytick distance = .0001,
    xminorticks = true,
    xlabel = {$t_i$, seconds},
    ylabel = {phasor angle (rad)},
    ylabel style   = {yshift = -.3em},
    x tick label style = {/pgf/number format/1000 sep={\,}},
    y tick label style = {/pgf/number format/1000 sep={\,}},
    cycle list name    = busPlotlist
  ]
  
    \foreach \y in {1,2,3}{
      \addplot+ table[x index = 0, y index = \y] {\tableINPUT};
    }    

        \node[anchor=north east, font=\small] at (rel axis cs:1,1.02) {Bus 61};

  \end{axis}
\end{tikzpicture}%
  \tikzexternaldisable%
\\
        \vspace{.5\baselineskip}
  \tikzexternalenable%
  \tikzsetnextfilename{singlesim_bus61errors_preEvent_thetaV}%
  \begin{tikzpicture}[font = \plotfontsize]
  \pgfplotstableread{graphics/data/single_sim_bus61errors_pre_thetaV.dat}\tableINPUT
  
  \begin{semilogyaxis}[%
    width  = .6\linewidth,
    height = .10\textheight,
    scale only axis,
    xmin = 5,
    xmax = 10,
    ymin = 1e-9,
    ymax = 1e2,
    /pgf/number format/fixed,
    /pgf/number format/precision=6,
    xminorticks = true,
    % yminorticks = true,
    xlabel = {$t_i$, seconds},
    ylabel = {relative error},
    ylabel style   = {yshift = -.3em},
    scaled x ticks = false,
    x tick label style = {/pgf/number format/1000 sep={\,}},
    y tick label style = {/pgf/number format/1000 sep={\,}},
    cycle list name    = busPlotlist
  ]

  \pgfplotsset{cycle list shift = 1}
    \foreach \y in {1,2}{
      \addplot+ table[x index = 0, y index = \y] {\tableINPUT};
    }    

    \addplot [domain=5:10, samples=2, solid, matlaborange, line width = 1.5pt] {0.3369};

    \addplot [domain=5:10, samples=2, solid, matlabgreen, line width = 1.5pt] {4.1213};
  \end{semilogyaxis}
\end{tikzpicture}%
  \tikzexternaldisable%
\\
        \vspace{\baselineskip}
    \end{subfigure}
    \begin{subfigure}[b]{.48\linewidth}
    \raggedleft
  \tikzexternalenable%
  \tikzsetnextfilename{singlesim_bus61reconstruction_during_thetaV}%
  \usetikzlibrary{spy}
\begin{tikzpicture}[font = \plotfontsize, spy using outlines={rectangle, magnification=4, size=1.5cm, connect spies}]
\pgfplotstableread{graphics/data/single_sim_bus61reconstruction_during_thetaV.dat}\tableINPUT
  
  \begin{axis}[%
    width  = .6\linewidth,
    height = .10\textheight,
    scale only axis,
    xmin = 29,
    xmax = 31,
    yticklabel style={/pgf/number format/.cd,fixed zerofill,precision=4},
    xminorticks = true,
    xlabel = {$t_i$, seconds},
    x tick label style = {/pgf/number format/1000 sep={\,}},
    y tick label style = {/pgf/number format/1000 sep={\,}},
    cycle list name    = busPlotlist
  ]

    \foreach \y in {1,2,3}{
      \addplot+ table[x index = 0, y index = \y] {\tableINPUT};
    }    

    \node[anchor=north east, font=\small] at (rel axis cs:1.03,1.02) {Bus 61};

  \end{axis}
\end{tikzpicture}%
  \tikzexternaldisable%
\\
        \vspace{.5\baselineskip}
  \tikzexternalenable%
  \tikzsetnextfilename{singlesim_bus61errors_during_thetaV}%
  \begin{tikzpicture}[font = \plotfontsize]
  \pgfplotstableread{graphics/data/single_sim_bus61errors_during_thetaV.dat}\tableINPUT
  
  \begin{semilogyaxis}[%
    width  = .6\linewidth,
    height = .10\textheight,
    scale only axis,
    xmin = 29,
    xmax = 31,
    ymin = 1e-9,
    ymax = 1e2,
    /pgf/number format/fixed,
    /pgf/number format/precision=6,
    xminorticks = true,
    % yminorticks = true,
    xlabel = {$t_i$, seconds},
    % ylabel = {$\relerr(t_i)$},
    % ylabel style   = {yshift = -.3em},
    scaled x ticks = false,
    x tick label style = {/pgf/number format/1000 sep={\,}},
    y tick label style = {/pgf/number format/1000 sep={\,}},
    cycle list name    = busPlotlist
  ]

  \pgfplotsset{cycle list shift = 1}
    \foreach \y in {1,2}{
      \addplot+ table[x index = 0, y index = \y] {\tableINPUT};
    }    

    \addplot [domain=29:31, samples=2, solid, matlaborange, line width = 1.5pt] {0.3369};

    \addplot [domain=29:31, samples=2, solid, matlabgreen, line width = 1.5pt] {4.1213};

  \node[overlay, text=black, font=\Large\bfseries] at (axis cs:30.02,0.3369) {$\times$};
  \end{semilogyaxis}
\end{tikzpicture}%
  \tikzexternaldisable%
\\
        \vspace{\baselineskip}
    \end{subfigure}\\
    \begin{subfigure}[b]{.48\linewidth}
    \raggedleft
  \tikzexternalenable%
  \tikzsetnextfilename{singlesim_bus68reconstruction_preEvent_thetaV}%
  \begin{tikzpicture}[font = \plotfontsize]
  \pgfplotstableread{graphics/data/single_sim_bus68reconstruction_pre_thetaV.dat}\tableINPUT
  
  \begin{axis}[%
    width  = .6\linewidth,
    height = .10\textheight,
    scale only axis,
    xmin = 5,
    xmax = 10,
    ymin = 0.7940,
    ymax = 0.8000,
    yticklabel style={/pgf/number format/.cd,fixed zerofill,precision=4},
    ytick = {0.7990,0.7970,0.7950},
    ytick distance = .0001,
    xminorticks = true,
    xlabel = {$t_i$, seconds},
    ylabel = {phasor angle (rad)},
    ylabel style   = {yshift = -.3em},
    x tick label style = {/pgf/number format/1000 sep={\,}},
    y tick label style = {/pgf/number format/1000 sep={\,}},
    cycle list name    = busPlotlist
  ]

    \foreach \y in {1,2,3}{
      \addplot+ table[x index = 0, y index = \y] {\tableINPUT};
    }    

    \node[anchor=north east, font=\small] at (rel axis cs:1.02,1.02) {Bus 68};
    
  \end{axis}
 \end{tikzpicture}%
  \tikzexternaldisable%
\\
        \vspace{.5\baselineskip}
  \tikzexternalenable%
  \tikzsetnextfilename{singlesim_bus68errors_preEvent_thetaV}%
  \begin{tikzpicture}[font = \plotfontsize]
  \pgfplotstableread{graphics/data/single_sim_bus68errors_pre_thetaV.dat}\tableINPUT
  
  \begin{semilogyaxis}[%
    width  = .6\linewidth,
    height = .10\textheight,
    scale only axis,
    xmin = 5,
    xmax = 10,
    ymin = 1e-9,
    ymax = 1e2,
    /pgf/number format/fixed,
    /pgf/number format/precision=6,
    xminorticks = true,
    % yminorticks = true,
    xlabel = {$t_i$, seconds},
    ylabel = {relative error},
    ylabel style   = {yshift = -.3em},
    scaled x ticks = false,
    x tick label style = {/pgf/number format/1000 sep={\,}},
    y tick label style = {/pgf/number format/1000 sep={\,}},
    cycle list name    = busPlotlist
  ]

  \pgfplotsset{cycle list shift = 1}
    \foreach \y in {1,2}{
      \addplot+ table[x index = 0, y index = \y] {\tableINPUT};
    }    

    \addplot [domain=5:10, samples=2, solid, matlaborange, line width = 1.5pt] {0.3369};

    \addplot [domain=5:10, samples=2, solid, matlabgreen, line width = 1.5pt] {4.1213};
  \end{semilogyaxis}
\end{tikzpicture}%
  \tikzexternaldisable%
\\
        \vspace{\baselineskip}
    \end{subfigure}
    \begin{subfigure}[b]{.48\linewidth}
    \raggedleft
  \tikzexternalenable%
  \tikzsetnextfilename{singlesim_bus68reconstruction_during_thetaV}%
  \begin{tikzpicture}[font = \plotfontsize]
  \pgfplotstableread{graphics/data/single_sim_bus68reconstruction_during_thetaV.dat}\tableINPUT
  
  \begin{axis}[%
    width  = .6\linewidth,
    height = .10\textheight,
    scale only axis,
    xmin = 29,
    xmax = 31,
    yticklabel style={/pgf/number format/.cd,fixed zerofill,precision=4},
    ytick = {0.9500,0.7500,0.5500},
    xminorticks = true,
    xlabel = {$t_i$, seconds},
    x tick label style = {/pgf/number format/1000 sep={\,}},
    y tick label style = {/pgf/number format/1000 sep={\,}},
    cycle list name    = busPlotlist,
  ]

    \foreach \y in {1,2,3}{
      \addplot+ table[x index = 0, y index = \y] {\tableINPUT};
    }    

    \node[anchor=north east, font=\small] at (rel axis cs:.95,1.02) {Bus 68};
    
  \end{axis}
\end{tikzpicture}%
  \tikzexternaldisable%
\\
        \vspace{.5\baselineskip}
  \tikzexternalenable%
  \tikzsetnextfilename{singlesim_bus68errors_during_thetaV}%
  \begin{tikzpicture}[font = \plotfontsize]
  \pgfplotstableread{graphics/data/single_sim_bus68errors_during_thetaV.dat}\tableINPUT
  
  \begin{semilogyaxis}[%
    width  = .6\linewidth,
    height = .10\textheight,
    scale only axis,
    xmin = 29,
    xmax = 31,
    ymin = 1e-9,
    ymax = 1e2,,
    /pgf/number format/fixed,
    /pgf/number format/precision=6,
    xminorticks = true,
    % yminorticks = true,
    xlabel = {$t_i$, seconds},
    % clip marker paths=true,
    % ylabel = {$\relerr(t_i)$},
    % ylabel style   = {yshift = -.3em},
    scaled x ticks = false,
    x tick label style = {/pgf/number format/1000 sep={\,}},
    y tick label style = {/pgf/number format/1000 sep={\,}},
    cycle list name    = busPlotlist
  ]

  \pgfplotsset{cycle list shift = 1}
    \foreach \y in {1,2}{
      \addplot+ table[x index = 0, y index = \y] {\tableINPUT};
    }    

    \addplot [domain=29:31, samples=2, solid, matlaborange, line width = 1.5pt] {0.3369};

    \addplot [domain=29:31, samples=2, solid, matlabgreen, line width = 1.5pt] {4.1213};
    
    % % \node[overlay, matlabred] at (axis cs:30.2,0.0424) {\Large$\times$};
    % \node[overlay, circle, fill=matlaborange, inner sep=2.5pt] at (axis cs:30.5,0.3369) {};
  \end{semilogyaxis}
\end{tikzpicture}%
  \tikzexternaldisable%
\\
        \vspace{\baselineskip}
    \end{subfigure}

    \vspace*{-\baselineskip}
    \caption{Interpolatory reconstruction and true data for the non-pilot voltage phasor angle datastreams at bus $61$ (top) and bus $68$ (bottom) before and during a three-phase fault of the line between buses $28$ and $29$. The black marker $\times$ indicates the precise instance at which the error estimator is violated according to~\eqref{eq:boundViolated} with $\theta = 1$.}
    \label{fig:singleSim_angle}
\end{figure}
%%%%%%%%%%%%%%%%%%%%%%%%%%%%%%%%%%%%%%%%%%%%%%%%%%%%%%%%%%%%%%%%%%%

Following training, the reconstruction errors for the non-pilot datastreams are monitored and compared against the values of the error indicator $\theta\,\etaBs\sigma_{K+1}$~\eqref{eq:boundViolated}.
For this test, we set $\theta = 1$ as a baseline.
The non-pilot voltage magnitude and phasor angle datastreams before and during the disturbance, along with their DEIM- and MILP-based reconstructions, are plotted in Figures~\ref{fig:singleSim_voltage} and~\ref{fig:singleSim_angle}. 
We also overlay the error plots with the values of the corresponding error estimator $\etaBs\sigma_{K+1}$.
Before the disturbance and during ambient operating conditions, both the DEIM- and MILP-based reconstructions of the non-pilot datastreams are accurate and well within their respective error estimates. 
Within one second of the line fault, the detection mechanism~\eqref{eq:boundViolated} is triggered by the DEIM-based reconstructions of the non-pilot voltage magnitude datastreams at both buses 63 and 67, and the non-pilot phasor angle datastream at bus 61.
Therefore, the fault is correctly detected.
For the non-pilot phasor angle datastream at bus $68$, the estimator is very nearly violated; this motivates rescaling the estimator by values of $\theta<1$.
We emphasize however that the disturbance is still correctly detected because the condition~\eqref{eq:boundViolated} is triggered by at least one of the non-pilot streams.
For the MILP-based reconstruction, the error remains within acceptable parameters according to the estimator (as $\etaBs$ is excessively large), and thus no fault is detected.

%%%%%%%%%%%%%%%%%%%%%%%%%%%%%%%%%%%%%%%%%%%%%%%%%%%%%%%%%%%%%%%%%%%
\begin{table}[t!]
  \caption{Precision, Recall, $\Fone$ and $\Ftwo$ scores for the event detection mechanism~\eqref{eq:boundViolated} using the DEIM- and MILP-based pilots for $\theta = 1$ and $10^{-2}$ applied to the $111$ three-phase line fault simulations.
  The largest scores in each row are highlighted in \textbf{boldface}.}
  \label{tab:scores_tp}
   \centering
  \vspace{.5\baselineskip}
  \begin{tabular}{lccccc}
    \hline\noalign{\smallskip}
     Voltage & DEIM & DEIM & MILP & MILP \\
      magnitudes & \footnotesize $\theta = 1$ 
      & \footnotesize $\theta = 10^{-2}$
      & \footnotesize $\theta = 1$
      & \footnotesize $\theta = 10^{-2}$\\
    \noalign{\smallskip}\hline\noalign{\smallskip}
    $\Prec$
      & \textbf{0.9515}
      & 0.9189
      & 0.9063
      & 0.9189 \\
    $\Rec$
      & 0.9245
      & \textbf{1.0000}
      & 0.2685
      & \textbf{1.0000} \\
    $\Fone$
      & 0.9378
      & \textbf{0.9577}
      & 0.4143
      & \textbf{0.9577} \\
    $\Ftwo$
      & 0.9298
      & \textbf{0.9827}
      & 0.3125
      & \textbf{0.9827} \\
    \noalign{\smallskip}\hline\noalign{\smallskip}
  \end{tabular}
  \vspace{\baselineskip}
    \begin{tabular}{lcccc}
    \hline\noalign{\smallskip}
     Phasor & DEIM & DEIM & MILP & MILP \\
      angles & \footnotesize $\theta = 1$ 
      & \footnotesize $\theta = 10^{-2}$
      & \footnotesize $\theta = 1$
      & \footnotesize $\theta = 10^{-2}$\\
    \noalign{\smallskip}\hline\noalign{\smallskip}
    $\Prec$
      & 0.3140
      & 0.9279
      & 0.0000
      & \textbf{0.9550} \\
    $\Rec$ 
      & 0.5192 
      & \textbf{1.0000}
      & 0.0000
      & \textbf{1.0000} \\
    $\Fone$
      & 0.3913
      & 0.9626
      & 0.0000
      & \textbf{0.9770} \\
    $\Ftwo$
      & 0.4592
      & 0.9847
      & 0.0000
      & \textbf{0.9907} \\
    \noalign{\smallskip}\hline\noalign{\smallskip}
  \end{tabular}
\end{table}
%%%%%%%%%%%%%%%%%%%%%%%%%%%%%%%%%%%%%%%%%%%%%%%%%%%%%%%%%%%%%%%%%%%

%%%%%%%%%%%%%%%%%%%%%%%%%%%%%%%%%%%%%%%%%%%%%%%%%%%%%%%%%%%%%%%%%%%
\begin{table}[t!]
  \caption{Precision, Recall, $\Fone$ and $\Ftwo$ scores for the event detection mechanism~\eqref{eq:boundViolated} using the DEIM- and MILP-based pilots for $\theta = 1$ and $10^{-2}$ applied to the $108$ line-to-ground fault simulations.
  The largest scores in each row are highlighted in \textbf{boldface}.}
  \label{tab:scores_tlg}
   \centering
  \vspace{.5\baselineskip}
  \begin{tabular}{lccccc}
    \hline\noalign{\smallskip}
     Voltage & DEIM & DEIM & MILP & MILP \\
      magnitudes & \footnotesize $\theta = 1$ 
      & \footnotesize $\theta = 10^{-2}$
      & \footnotesize $\theta = 1$
      & \footnotesize $\theta = 10^{-2}$\\
    \noalign{\smallskip}\hline\noalign{\smallskip}
    $\Prec$
      & 0.9600
      & 0.9537
      & \textbf{1.0000}
      & 0.9537 \\
    $\Rec$
      & 0.9231
      & \textbf{1.0000}
      & 0.2870
      & \textbf{1.0000} \\
    $\Fone$
      & 0.9412
      & \textbf{0.9763}
      & 0.4460
      & \textbf{0.9763} \\
    $\Ftwo$
      & 0.9302
      & \textbf{0.9904}
      & 0.3348
      & \textbf{0.9904} \\
    \noalign{\smallskip}\hline\noalign{\smallskip}
  \end{tabular}
  \vspace{\baselineskip}
    \begin{tabular}{lcccc}
    \hline\noalign{\smallskip}
     Phasor & DEIM & DEIM & MILP & MILP \\
      angles & \footnotesize $\theta = 1$ 
      & \footnotesize $\theta = 10^{-2}$
      & \footnotesize $\theta = 1$
      & \footnotesize $\theta = 10^{-2}$\\
    \noalign{\smallskip}\hline\noalign{\smallskip}
    $\Prec$
      & 0.3059
      & 0.9630
      & 0.0000
      & \textbf{0.9907} \\
    $\Rec$ 
      & 0.5306
      & \textbf{1.0000}
      & 0.0000
      & \textbf{1.000} \\
    $\Fone$
      & 0.3881
      & 0.9811
      & 0.0000
      & \textbf{0.9953} \\
    $\Ftwo$
      & 0.4626
      & 0.9923
      & 0.0000
      & \textbf{0.9981} \\
    \noalign{\smallskip}\hline\noalign{\smallskip}
  \end{tabular}
\end{table}
%%%%%%%%%%%%%%%%%%%%%%%%%%%%%%%%%%%%%%%%%%%%%%%%%%%%%%%%%%%%%%%%%%%

%%%%%%%%%%%%%%%%%%%%%%%%%%%%%%%%%%%%%%%%%%%%%%%%%%%%%%%%%%%%%%%%%%
\begin{table}[t]
  \caption{Precision, Recall, $\Fone$ and $\Ftwo$ scores for the event detection mechanism~\eqref{eq:boundViolated} using the DEIM- and MILP-based pilots for $\theta = 1$ and $10^{-2}$ applied to the $140$ loss-of-line (tripping) simulations.
  The largest scores in each row are highlighted in \textbf{boldface}.}
  \label{tab:scores_lineloss}
   \centering
  \vspace{.5\baselineskip}
  \begin{tabular}{lccccc}
    \hline\noalign{\smallskip}
     Voltage & DEIM & DEIM & MILP & MILP \\
      magnitudes & \footnotesize $\theta = 1$ 
      & \footnotesize $\theta = 10^{-2}$
      & \footnotesize $\theta = 1$
      & \footnotesize $\theta = 10^{-2}$\\
    \noalign{\smallskip}\hline\noalign{\smallskip}
    $\Prec$
      & 0.2667
      & \textbf{0.9542}
      & 0.0000
      & 0.9189 \\
    $\Rec$ 
      & 0.0310
      & \textbf{0.9328}
      & 0.0000
      & 0.5075\\
    $\Fone$
      & 0.0556
      & \textbf{0.9434}
      & 0.0000
      & 0.6538 \\
    $\Ftwo$
      & 0.0377
      & \textbf{0.9370}
      & 0.0000
      & 0.5574 \\
    \noalign{\smallskip}\hline\noalign{\smallskip}
  \end{tabular}
  \vspace{\baselineskip}
    \begin{tabular}{lcccc}
    \hline\noalign{\smallskip}
     Phasor & DEIM & DEIM & MILP & MILP \\
      angles & \footnotesize $\theta = 1$ 
      & \footnotesize $\theta = 10^{-2}$
      & \footnotesize $\theta = 1$
      & \footnotesize $\theta = 10^{-2}$\\
    \noalign{\smallskip}\hline\noalign{\smallskip}
    $\Prec$
      & 0.1667
      & \textbf{0.9516}
      & 0.0000
      & 0.8901 \\
    $\Rec$
      & 0.0667
      & \textbf{0.8806}
      & 0.0000
      & 0.6231 \\
    $\Fone$
      & 0.0952
      & \textbf{0.9147}
      & 0.0000
      & 0.7330 \\
    $\Ftwo$
      & 0.0758
      & \textbf{0.8939}
      & 0.0000
      & 0.6628 \\
    \noalign{\smallskip}\hline\noalign{\smallskip}
  \end{tabular}
\end{table}
%%%%%%%%%%%%%%%%%%%%%%%%%%%%%%%%%%%%%%%%%%%%%%%%%%%%%%%%%%%%%%%%%%%

For the second test, we verify the robustness of the approach in~\eqref{eq:boundViolated} in detecting disturbances on a collection of $359$ event simulation scenarios of the 68-bus, 16-machine test system in MATLAB's PST. 
The dataset comprises three-phase line faults ($111$), line-to-ground faults ($108$), and loss-of-line events (tripping) without an electrical fault ($140$). The scenarios were generated by cycling through all transmission lines and applying the disturbance.
The length of each simulation window is $\To=100$\,s.
Each simulation is initialized from the same pre-disturbance operating point obtained by solving the network load-flow equations and initializing all dynamic states consistently with that operating point. For each disturbed line, we generate two distinct event scenarios: in the first, the disturbance occurs at a randomly selected time in the first half of the 100 s simulation window, and in the second, the disturbance occurs at a randomly selected time in the second half of the window. 
For a subset of these simulations, the inner Newton solve for the nonlinear-load bus voltages did not converge following the fault; these were not included in the final collection of $359$ simulations.
As before, the data are voltage magnitudes and phasor angles at each bus sampled at a rate of $100$\,Hz.

For monitoring the voltage magnitude and phasor angle datastreams, we use the same DEIM- and MILP-based pilot configurations reported in Tables~\ref{tab:pilots_voltage} and~\ref{tab:pilots_angle}.
We also monitor the same non-pilot voltage magnitude datastreams at buses $63,67$ and non-pilot phasor angle datastreams at buses $61,68$, as in the previous test.
For each simulation scenario, we apply the detection mechanism~\eqref{eq:boundViolated} \emph{separately} to the monitored voltage magnitude and phasor angle datastreams, and assess the resulting decisions independently.
For each data type and simulation scenario, we classify the decision made by~\eqref{eq:boundViolated} as a $\TP$ (true positive: the disturbance is correctly detected at one of the non-pilots within $1\,$s of the event according to~\eqref{eq:boundViolated}), an $\FP$ (a fault is detected outside of this window at both or one of the non-pilots, and not detected correctly at the other), or an $\FN$ (no fault is detected at either of the non-pilots).
We then evaluate the performance of the detection mechanism~\eqref{eq:boundViolated} using $\Fone$ and $\Ftwo$ scores~\cite{GouG05}, computed as
\begin{equation*}
    \Fone = \frac{2\times \Prec \times \Rec}{\Prec + \Rec},~\Ftwo = \frac{5\times \Prec \times \Rec}{4 \times \Prec + \Rec},
\end{equation*}
for $\Prec = \TP/(\TP + \FP)$, $\Rec = \TP/(\TP + \FN)$.
Values close to $1$ indicate a reliable detection mechanism.
If $\Prec$ and $\Rec$ are both zero, we report the corresponding $\Fone$ and $\Ftwo$ scores as zero.

For the voltage magnitude and phasor angle data, we compute the $\Fone$ and $\Ftwo$ scores for four different detection mechanisms: \eqref{eq:boundViolated} using the DEIM-based pilots with $\theta = 1$ and $\theta = 10^{-2}$, and~\eqref{eq:boundViolated} using the MILP-based pilots with the same two values of $\theta$. 
We compute these scores separately for the three types of event simulations contained in the dataset: three-phase line faults, line-to-ground faults, and line tripping without a fault.
These scores, along with the associated $\Prec$ and $\Rec$, are recorded in Tables~\ref{tab:scores_tp},~\ref{tab:scores_tlg}, and~\ref{tab:scores_lineloss}.
These experiments investigate the fidelity of the detection mechanism~\eqref{eq:boundViolated} as it applies to different types of disturbances.

We generally observe a substantial improvement in detection when $\theta$ is decreased from $1$ to $10^{-2}$. 
For the MILP-based pilot configuration, this improvement is extremely significant across the board,
but most noticeably for the loss-of-line events.  We attribute this behavior to the fact that the data deviate less dramatically from nominal operating conditions in response to such an event, and hence the unscaled estimator corresponding to $\theta = 1$ is less effective at capturing these small-scale deviations.
For the voltage magnitude data, the DEIM-based pilots significantly outperform the MILP-based pilots for $\theta = 1$, and are on par with the MILP-based pilots for $\theta = 10^{-2}$.
For the phasor angle data, the DEIM-based pilots perform better for $\theta = 1$, whereas the MILP-based pilots perform better for $\theta = 10^{-2}$.

%%%%%%%%%%%%%%%%%%%%%%%%%%%%%%%%%%%%%%%%%%%%%%%%%%%%%%%%%%%%%%%%%%%
\subsection{Event Localization Using DEIM}
%%%%%%%%%%%%%%%%%%%%%%%%%%%%%%%%%%%%%%%%%%%%%%%%%%%%%%%%%%%%%%%%%%%
After a disturbance occurs, it is imperative to find its source, e.g., the buses adjacent to a faulted line, quickly, so system operators can take corrective action to prevent cascading failures.
Numerous works have explored the event location problem; see, e.g.~\cite{LiWC18,WanZZ11,Liuetal15}.
As an alternative to these approaches, we propose using the DEIM algorithm to localize the source of disturbances.
Once a disturbance has been detected using~\eqref{eq:boundViolated} (or any other detection mechanism),
DEIM can be applied to a batch of data from \emph{all} datastreams containing the transient system response following the disturbance. 
In our tests, as little as $1$\,s of data following the disturbance is needed to localize the event.

%%%%%%%%%%%%%%%%%%%%%%%%%%%%%%%%%%%%%%%%%%%%%%%%%%%%%%%%%%%%%%%%%%%
\emph{Numerical Tests.} 
%%%%%%%%%%%%%%%%%%%%%%%%%%%%%%%%%%%%%%%%%%%%%%%%%%%%%%%%%%%%%%%%%%%
We demonstrate the ability of DEIM to localize disturbances using the same 359 simulation scenarios from Section~\ref{ss:eventDet}.
In the interest of space, we only report results for the voltage magnitude data. Results for the phasor angle data are available in the accompanying code package~\cite{supRei25}.
For comparison, we use the data-driven energy-based (EB) criterion for localizing affected buses from Section III.D of~\cite{LiWC18}.
For each scenario, we assume that an event alert has been correctly issued. We then aggregate $0.5\,$s of data prior to the event and $1.0\,s$ of data directly after the event into the matrix $\BY$.\ \ 
Before attempting to localize the event's source, this matrix is preprocessed by removing the mean of the pre-event data, as in~\cite{LiWC18}.
We apply DEIM, QDEIM, and EB to the leading $K$ left singular vectors $\BU_K$ of $\BY$ to select up to $K=1,2,\ldots, 10$ rows, each of which corresponds to a particular PMU in the network.
Let $\CE_{K}$ denote the top $K$ buses identified by a given method, e.g., DEIM.
Let $\CB_{i}:=\{s_{i_{1}},s_{i_{2}}\}\subseteq \{1,\ldots, N\}$ contain the indices corresponding to the source of event $i$ in the dataset; that is, $\CB_{i}$ is a set containing a pair of bus indices connecting a faulted line for the line-based events.
We measure the success of our method via the accuracy score
\begin{equation}
    \label{eq:eventLocAcc}
    \acc(K)=\frac{1}{\Ne} \sum_{i=1}^{\Ne} 1(\CB_{i} \subseteq \CE_{K})\times 100 \,\%,
\end{equation}
where $\Ne\geq 0$ is the number of events in the dataset, and $ 1(\CB_{i} \subseteq \CE_{K})$ is an indicator function that equals $1$ if $\CB_{i} \subseteq \CE_{K}$, i.e., the method correctly captures the source of the disturbance in its $K$ candidate locations, and $0$ otherwise.
If \emph{both} buses connected to the affected line are found within these $k$ indices, we classify the method as having correctly localized the source of the event. 
This process is repeated for each of the $359$ event simulation scenarios.

%%%%%%%%%%%%
\begin{figure}[t!]
\centering
  \tikzexternalenable%
  \tikzsetnextfilename{eventLegend}%
  \begin{tikzpicture}
  \begin{axis}[%
    hide axis,
    width  = 1mm,
    height = 1mm,
    scale only axis,
    xmin = 0,
    xmax = 1,
    ymin = 0,
    ymax = 1,
    legend columns = 3, 
    legend style   = {
      at     = {(0,0)},
      anchor = center,
      /tikz/every even column/.append style = {column sep = 0.2cm}},
    legend cell align  = {left},
    clip mode          = individual,
    cycle list name    = interpApproxPlotlist]

    \foreach \y in {1, 2, 3}{
      \addplot+ coordinates{ (0, 0) };
    }
    \plotfontsize
    \addlegendentry{DEIM}
    \addlegendentry{QDEIM}
    \addlegendentry{EB}
    % \addlegendentry{...}
  \end{axis}
\end{tikzpicture}%
  \tikzexternaldisable%

\vspace*{0.5\baselineskip}
\begin{subfigure}[b!]{\linewidth}
    \centering
  \tikzexternalenable%
  \tikzsetnextfilename{event_loc_tp_busV}%
  \begin{tikzpicture}[font = \plotfontsize]
  \pgfplotstableread{graphics/data/event_loc_tp_busV.dat}\tableINPUT
  
  \begin{axis}[%
    width  = .75\linewidth,
    height = .1\textheight,
    scale only axis,
    grid=both,
    grid style={line width=.1pt, draw=gray!10},
    major grid style={line width=.2pt,draw=gray!25},
    minor tick num=1,
    xmin = 2,
    xmax = 10,
    ymin = 0,
    ymax = 100,
    xminorticks = true,
    yminorticks = true,
    xlabel = {$K$, number of candidate locations},
    ylabel = {$\acc(K)$},
    ylabel style   = {yshift = -.3em},
    scaled x ticks = false,
    x tick label style = {/pgf/number format/1000 sep={\,}},
    y tick label style = {/pgf/number format/1000 sep={\,}},
    cycle list name    = interpApproxPlotlist
  ]
  
    \foreach \y in {1,2,3}{
      \addplot+ [restrict x to domain=2:10]  table[x index = 0, y index = \y] {\tableINPUT};
    }
    
  \end{axis}
\end{tikzpicture}%
  \tikzexternaldisable%

    \caption{Localization accuracy of three-phase line fault event simulations (108 scenarios) for $K=2,\ldots,10$.}
    \label{fig:per_correct_tp}
    \vspace{.5\baselineskip}
\end{subfigure}
\begin{subfigure}[b!]{\linewidth}
    \centering
  \tikzexternalenable%
  \tikzsetnextfilename{event_loc_ltg_busV}%
  \begin{tikzpicture}[font = \plotfontsize]
  \pgfplotstableread{graphics/data/event_loc_ltg_busV.dat}\tableINPUT
  
  \begin{axis}[%
    width  = .75\linewidth,
    height = .1\textheight,
    scale only axis,
    grid=both,
    grid style={line width=.1pt, draw=gray!10},
    major grid style={line width=.2pt,draw=gray!25},
    minor tick num=1,
    xmin = 2,
    xmax = 10,
    ymin = 0,
    ymax = 100,
    xminorticks = true,
    yminorticks = true,
    xlabel = {$K$, number of candidate locations},
    ylabel = {$\acc(K)$},
    ylabel style   = {yshift = -.3em},
    scaled x ticks = false,
    x tick label style = {/pgf/number format/1000 sep={\,}},
    y tick label style = {/pgf/number format/1000 sep={\,}},
    cycle list name    = interpApproxPlotlist
  ]
  
    \foreach \y in {1,2,3}{
      \addplot+ [restrict x to domain=2:10]  table[x index = 0, y index = \y] {\tableINPUT};
    }
    
  \end{axis}
\end{tikzpicture}%
  \tikzexternaldisable%

    \caption{Localization accuracy of line-to-ground line fault event simulations (108 scenarios) for $K=2,\ldots,10$.}
    \label{fig:per_correct_ltg}
    \vspace{.5\baselineskip}
\end{subfigure}
\begin{subfigure}[b!]{\linewidth}
    \centering
  \tikzexternalenable%
  \tikzsetnextfilename{event_loc_lossofline_busV}%
  \begin{tikzpicture}[font = \plotfontsize]
  \pgfplotstableread{graphics/data/event_loc_lossofline_busV.dat}\tableINPUT
  
  \begin{axis}[%
    width  = .75\linewidth,
    height = .1\textheight,
    scale only axis,
    grid=both,
    grid style={line width=.1pt, draw=gray!10},
    major grid style={line width=.2pt,draw=gray!25},
    minor tick num=1,
    xmin = 2,
    xmax = 10,
    ymin = 0,
    ymax = 100,
    xminorticks = true,
    yminorticks = true,
    xlabel = {$K$, number of candidate locations},
    ylabel = {$\acc(K)$},
    ylabel style   = {yshift = -.3em},
    scaled x ticks = false,
    x tick label style = {/pgf/number format/1000 sep={\,}},
    y tick label style = {/pgf/number format/1000 sep={\,}},
    cycle list name    = interpApproxPlotlist
  ]
  
    \foreach \y in {1,2,3}{
      \addplot+ [restrict x to domain=2:10]  table[x index = 0, y index = \y] {\tableINPUT};
    }
    
  \end{axis}
\end{tikzpicture}%
  \tikzexternaldisable%

    \caption{Localization accuracy of loss-of-line event simulations (108 scenarios) for $K=2,\ldots,10$.}
    \label{fig:per_correct_lossofline}
    \vspace{.5\baselineskip}
\end{subfigure}

\caption{\label{fig:per_correct}
Localization accuracy of all event simulations by DEIM, QDEIM, and EB for $K=2,\ldots,10$ using voltage magnitude data.}
\end{figure}
%%%%%%%%%%%%

The accuracies~\eqref{eq:eventLocAcc} with which DEIM, QDEIM, or EB correctly identified the source of the event within $K=2,\ldots,10$ indices are plotted in Figure~\ref{fig:per_correct}.
For the three-phase and line-to-ground fault events and values of $K\geq 4$, all methods localize the source of the faulted line with greater than $90$\,percent accuracy.
Significantly, DEIM and QDEIM are able to identify the source of the fault with 100\% accuracy for $K \geq 5$.
Thus, given the freedom to select enough indices, our DEIM- and QDEIM-based localization strategies correctly identify both affiliate buses connected to the faulted line in these fault-based scenarios, and perform marginally better than the reference approach in~\cite{LiWC18}.
For the loss-of-line events without a fault, none of the considered methods achieve an accuracy score greater than $40$\,percent, and EB performs best for all values of $K=2,\ldots,10$.
We observed similar behavior for the phasor angle data. In general, the source of the disturbance was harder to localize from these data, although EB was the most reliable on average.

%%%%%%%%%%%%%%%%%%%%%%%%%%%%%%%%%%%%%%%%%%%%%%%%%%%%%%%%%%%%%%%%%%%%%%%%%%%%%%%%
% CONCLUSIONS.                                                                 %
%%%%%%%%%%%%%%%%%%%%%%%%%%%%%%%%%%%%%%%%%%%%%%%%%%%%%%%%%%%%%%%%%%%%%%%%%%%%%%%%

%%%%%%%%%%%%%%%%%%%%%%%%%%%%%%%%%%%%%%%%%%%%%%%%%%%%%%%%%%%%%%%%%%%%%%%%%%%%%%%%
\section{Conclusions} \label{sec:conc}
%%%%%%%%%%%%%%%%%%%%%%%%%%%%%%%%%%%%%%%%%%%%%%%%%%%%%%%%%%%%%%%%%%%%%%%%%%%%%%%%
Interpolatory matrix decompositions (IDs) and the discrete empirical interpolation method (DEIM) provide effective tools for PMU data compression, network monitoring, and event detection.
We have shown that IDs give effective low-rank approximations of PMU data, particularly for time-sensitive and bandwidth-limited applications such as wide-area monitoring. IDs can be maintained in real-time while interacting with $K\ll N$ pilot streams or $K\ll T$ pilot snapshots.
Casting data compression in the mathematical framework of IDs provides a rigorous upper bound on the training error. 
To identify pilot buses for online monitoring, we employ DEIM, a greedy method that yields favorable error bounds.
DEIM can be applied during offline training to adaptively select pilots until the error bound falls below a user's tolerance, yielding an estimate of the online reconstruction error.
Any significant deterioration in this error estimate signals a notable change in the network's operating status relative to training conditions, providing a mechanism to detect disturbances.
We have shown that DEIM and its QDEIM variant can robustly localize the source of disturbances.

\appendix
This appendix describes the Mixed-Integer Linear Programming (MILP) formulation discussed in Section~\ref{ss:interpNumerics}. Given a data matrix $\BY\in\R^{N\times T}$, the goal is to select a subset of $K$ rows that are most orthogonal with each other. We next formally define the orthogonality metric. Following the procedure introduced in~\cite{XieCK14}, the matrix $\BY$ is first projected onto its $m \geq K$ leading principal components, where $m$ is chosen so that a certain amount of variance is preserved. Let $\wtBY\in\R^{N\times T}$ denote the projected matrix, and $\wtBY_{i,:}^{\trans}$ denote its $i$-th row. If $\vartheta_{i,j}$ is the angle between the vectors $\wtBY_{i,:}^{\trans}$ and $\wtBY_{j,:}^{\trans}\in\R^{1\times T}$, let us define the \emph{cosine similarity}:
\begin{equation}
c_{i,j}:=\cos(\vartheta_{i,j})=\frac{\wtBY_{i,:}\wtBY_{j,:}^{\trans}}{\|\wtBY_{i,:}\|_{2}\cdot \|\wtBY_{j,:}\|_{2}}.
\end{equation}
Note that $c_{i,j}$ is a scalar because $\wtBY_{i,:}\wtBY_{j,:}^{\trans}$ is an inner product. 
Reference~\cite{XieCK14} selects a subset $\mathcal{S}$ of $K$ rows so that the cosine similarity among the selected datastreams is as close to zero as possible (reflecting near-orthogonality of the vectors).
We formulate this goal as the subset selection problem
\begin{align}\label{eq:cosOpt}
\min_{\substack{\CS\subseteq \{1,\ldots,N\}\\|\CS|=K}}\max_{\substack{i<j\\i,j\in\CS}} |c_{i,j}|.
\end{align}
Solving~\eqref{eq:cosOpt} amounts to choosing $K$ rows of $\wtBY$ such that the largest pairwise (unsigned) cosine similarity among the selected rows is as small as possible, and hence the selected rows are as close to pairwise orthogonal as possible.

Problem~\eqref{eq:cosOpt} is not straightforward to solve, but can be posed as a MILP.\ \ For each row, introduce a binary decision variable taking the value $z_i=1$ if the $i$-th row is selected, and $z_i=0$ otherwise.  Problem~\eqref{eq:cosOpt} can be reformulated as:
\begin{subequations}\label{eq:milpOpt}
\begin{align}
\min_{x,\{z_i\}_{i=1}^N}~&~x\label{eq:milpOpt:a}\\
\mathrm{subject~to}~&~|c_{i,j}|(z_i+z_j-1)\leq x,~~\forall i<j,\label{eq:milpOpt:b}\\
&~\sum_{i=1}^N z_i=K,\label{eq:milpOpt:c}\\
&~x\geq 0,~~ z_i\in\{0,1\}~~\forall i.\label{eq:milpOpt:d}
\end{align}
\end{subequations}
Because the epigraph variable $x$ satisfies $x\geq 0$, the constraint~\eqref{eq:milpOpt:b} is non-redundant only if $z_i=z_j=1$. For those pairs, the constraint becomes $x\geq |c_{i,j}|$. Compiling all those pairs corresponding to non-redundant constraints in~\eqref{eq:milpOpt:b}, we get that $x\geq \max_{i,j\in\mathcal{S}}|c_{i,j}|$. Constraint~\eqref{eq:milpOpt:c} enforces a budget of $K$ rows. We solved this MILP in MATLAB using the \texttt{intlinprog} command with default settings. If instead of $K$ rows, we want to select $K$ columns of a matrix, the previous method would be applied to the transpose of $\BY$.
% %%%%%%%%%%%%%%%%%%%%%%%%%%%%%%%%%%%%%%%%%%%%%%%%%%%%%%%%%%%%%%%%%%%%%%%%%%%%%%%

%%%%%%%%%%%%%%%%%%%%%%%%%%%%%%%%%%%%%%%%%%%%%%%%%%%%%%%%%%%%%%%%%%%%%%%%%%%%%%%%
% \section*{Acknowledgments}
%%%%%%%%%%%%%%%%%%%%%%%%%%%%%%%%%%%%%%%%%%%%%%%%%%%%%%%%%%%%%%%%%%%%%%%%%%%%%%%%
\balance
\bibliographystyle{IEEEtran}
\bibliography{main}

\end{document}